\documentstyle[11pt,dina4,epsf]{article}
\renewcommand{\baselinestretch}{1.3}
\begin{document}
\pagestyle{empty}
\renewcommand{\baselinestretch}{1.3}
\renewcommand{\theequation}{\arabic{section}.\arabic{equation}}
\begin{flushright}
MS--TPI--94--13\\
December 1994\\
hep-th/9502157
\mbox{}\\
\mbox{}\\
\mbox{}\\
\end{flushright}
\begin{center}
{\Large \bf{A Systematic Extended Iterative Solution}}\\
{\Large \bf{for Quantum Chromodynamics
}}
\end{center}
\mbox{}\\
\mbox{}\\
\mbox{}\\
\begin{center}
M. Stingl\\
Institute for Theoretical Physics I, University of M\"unster\\
D-48149 M\"unster, Germany
\end{center}
\mbox{}\\
\mbox{}\\
\mbox{}\\
\mbox{}\\
{\underline{Abstract:}} An outline is given of an extended perturbative
solution of Euclidean QCD which systematically accounts for a class of
nonperturbative effects, while still allowing renormalization by the
perturbative counterterms. Euclidean proper vertices $\Gamma$ are
approximated by a double sequence $\Gamma^{[r,p]}$, where $r$ denotes
the degree of rational approximation with respect to the spontaneous
mass scale $\Lambda_{QCD}$, nonanalytic in the coupling $g^{2}$, while
$p$ represents the order of perturbative corrections in $g^{2}$
calculated from $\Gamma^{[r,0]}$ -- rather than from the perturbative
Feynman rules $\Gamma^{(0) {\rm{pert}}}$ -- as a starting point.
The mechanism allowing the nonperturbative
terms to reproduce themselves in the Dyson-Schwinger equations
preserves perturbative renormalizability and
is intimately tied to the divergence structure of the theory. As a
result,
it restricts the self-consistency problem for the $\Gamma^{[r,0]}$
rigorously -- i.e. without decoupling approximations -- to the seven
superficially divergent vertices. An interesting
aspect of the solution is that rational-function sequences for the QCD
propagators contain subsequences describing short-lived elementary
excitations. The method is calculational, in that it allows the known
techniques of loop computation to be used while dealing with integrands
of truly nonperturbative content.
\newpage
\pagestyle{plain}
\setcounter{page}{1}
\section{Generalities and Notation}
\subsection{Nonperturbative Quantities}
\setcounter{equation}{0}
 
One of the more important insights to have emerged from two decades
of study of the large-order behavior and summability of perturbation
expansions [1] has been that for renormalizable but not
superrenormalizable field theories (typically in four dimensions), the
perturbative series of
correlation functions $\Gamma (g^{2})$ around $g^{2} = 0$ is
fundamentally incomplete, in the sense that it does not allow unique
reconstruction of those functions even in principle. Perturbation
series, in most cases of interest, are divergent asymptotic series which
at fixed positive $g^{2}$ allow a function to be estimated with a finite
accuracy. A typical error estimate is
\begin{equation}
\mathrel{\mathop {\rm min}_{\scriptstyle \{p\}}} \, \, \left| \,
\, \Gamma (g^{2}) - \sum
\limits^{p}_{p' = 0} \, \,
\Gamma^{(p') {\rm{pert}}} \left( \frac{g}{4 \pi} \right)^{2p'} \, \,
\right|\, \, \propto \, \,
e^{-\frac{\rm const.}{g^{2}}} \, ,
\end{equation}
as can be inferred from the behavior of the perturbative coefficients
$\Gamma^{(p){\rm{pert}}}$
generally found in quantum field theories [1]. For superrenormalizable
theories, typically in dimensions $D \le 3$, analyticity properties of
the $\Gamma's$ in the complex $g^{2}$ plane can be established outside
of perturbation theory that
are strong enough to conclude that a function sufficiently analytic and
satisfying this bound must in fact vanish, and that $\Gamma's$ can be
reconstructed uniquely from their asymptotic series by resummation
techniques. By contrast, for renormalizable theories in $D = 4$, the
necessary analyticity is positively known to be violated [2], and
$\Gamma's$ may therefore differ from their perturbative series by terms
exponentially small near $g^{2} = 0$ that can "duck under" the bound
(1.1) and remain invisible in the expansion.
 
The general mathematics of semi-convergent series does not say more
about the missing terms, if any, that have no expansion in $g^{2}$ in
even a formal sense : one must turn to physics for clues. Asymptotically
free theories are special in that there one has a-priori knowledge, from
renormalization-group (RG) analysis, of the presence of one important
class of terms just allowed by the bound (1.1). These are the terms
involving the RG-invariant spontaneous mass scale [3],
\begin{equation}
\Lambda^{2} \left(g^{2}(\nu), \nu \right)_{R} =
\nu^{2} \exp \left\{ -2 \int \limits^{g (\nu)} \, \,
d g' \, \frac{1}{[\beta (g')]_{R}} \right\} \, ,
\end{equation}
where $\nu$ is the arbitrary mass scale introduced by
renormalization in a scheme $R$. With asymptotic freedom,
\begin{equation}
\Lambda^{2} = \nu^{2} \exp \left\{ - \frac{1}
{\beta_{0} [g(\nu)/4 \pi]^{2}} [1 + O(g^{2})] \right\} \, ,
\end{equation}
(where $\beta_{0}$, defined by eq. (1.20) below, is positive), the
archetypal non-analytic
$g^{2}$ dependence
of the "exponentially small" form.
It hardly needs emphasis that the term "exponentially small" refers
to the hypothetical behavior of such a quantity in a purely formal,
unphysical limit -- the limit where one sends $g^{2} \rightarrow 0$
{\it{without}} a corresponding increase in $\nu$ in the sense of $RG$
flow. In the commonly used (dimensional) schemes $R, \, \Lambda_{R}$ of
course is not small, and in fact sets the scale of hadronic masses.
 
It is also clear from the context that
when speaking of $\Lambda^{2}$-dependent terms
we are {\it{not}} alluding here to the relatively weak $\Lambda^{2}$
dependence obtained by the standard RG process of introducing a
running coupling constant
$\bar{\alpha} (k^{2}/\Lambda^{2})$, with asymptotic expansion
\begin{equation}
\frac{\bar{\alpha} (z)}{4 \pi} = \frac{1}{\beta_{0} \ln z} \left\{
1 + O \left( \frac{1}{\ln z} \right) \right\} \qquad
\left( z = \frac{k^{2}}{\Lambda^{2}} >\!> 1 \right) \, ,
\end{equation}
since this arises purely {\it{within}} the perturbation series by
leading-logarithms $(LL)$ resummation. By contrast, consider some
quantities presumed to be nonperturbative in the deeper sense mentioned
above, such as the vacuum expectation of the trace anomaly [4] of the
energy-momentum tensor in massless QCD. Since this is known to be an RG
invariant and of mass dimension four, it must be of the form
\begin{equation}
<0 \left| \frac{\beta(g)}{2 g} G^{\mu \nu}_{a} \, G^{\mu \nu}_{a}
\right|
0>_{R}
= c_{A} \Lambda^{4} \, ,
\end{equation}
where $c_{A}$ is a pure number. As another example, again in massless
QCD, consider dynamical formation of a color-singlet glueball state,
whose mass defines an RG-invariant scale. The connected and amputated
four-gluon amplitude takes the form,
\begin{equation}
T^{\kappa \lambda \mu \nu}_{abcd} (p_{1} \ldots p_{4}) = \delta_{ab}
\delta_{cd} \frac{\Phi^{\kappa \lambda} (p_{1}, p_{2}) \Phi^{\mu \nu}
(p_{3}, p_{4})}{(p_{1} + p_{2})^{2} + c_{B} \Lambda^{2}}
\end{equation}
\begin{displaymath}
{\rm { + (crossed \, \, \, terms) + (regular \, \, \, terms)}} \, ,
\end{displaymath}
with $c_{B}$ again a pure number. Yet another example is furnished by
instanton effects.
There, $n$-instanton contributions to physical quantities typically
appear with factors $\exp (-n S_{1}/g^{2})$, where the instanton action
$S_{1}$ is in general {\it{not}} an integer multiple of the $(4
\pi)^{2}/\beta_{0}$ of eq. (1.3), so that fractional powers of $\Lambda$
seem to be present. However, strong arguments have been given by
M\"unster [5] to the effect that this is really an artefact of the
dilute-gas approximation used in most instanton calculations,
and that by taking account of the denseness of the instanton system one
is led back to integer powers of $\Lambda$ in the properly summed
multi-instanton series for physical quantities. These examples indicate,
as does experience with operator-product expansions, that the $\Lambda$
dependence of truly nonperturbative quantities will typically be of a
{\it{polynomial or rational}} form. It is this type of dependence which
can be clearly separated, without double-counting problems, from the
"resummed-perturbative" one of (1.4), and for which $g^{2} \, \, and \,
\, \Lambda^{2}$ {\it{can be handled formally like two different
parameters,}} (although in the end one will always seek to use
$LL$ resummation to eliminate $g^{2}$ and deal only with a single,
$RG$-invariant parameter, $\Lambda^{2}$).
 
The present paper outlines [6], for an Euclidean and asymptotically free
gauge theory exemplified by QCD, a systematic extension of perturbation
theory designed to account for {\it{nonperturbative effects as reflected
in a rational dependence, or more generally a dependence representable
by sequences of rational functions,}} of correlation functions on
$\Lambda$.
Since the approximating sequence is to provide a genuine extension,
rather than a resummation, of the perturbative one, it is natural that
it should take the form of a {\it{double sequence}}, $\Gamma^{[r,p]}$,
as described in sect. 2 for the simplest case of a one-component vertex
function depending on a single squared Euclidean momentum (the two-point
vertex of transverse gluons); sect. 3 and appendix A describe the
relatively
straightforward extension to the more complex (three- and four-point)
basic vertex functions. In this double sequence, the integer $r$
will characterize a certain degree of approximation with respect to
$\Lambda$, while the "perturbative" index $p$ still counts explicit
powers of $g^{2}(\nu)$. That $r$ cannot simply be an index counting
powers of $\Lambda$ has to do with the fundamental difference of
dimensionality: with $\Lambda$ being a mass, as opposed to the
dimensionless $g^{2}$, an expansion in powers of $\Lambda$ would (for
fixed mass dimension $d_{\Gamma}$ of $\Gamma$) inevitably be a
large-$Q^{2}$ expansion in powers of $\Lambda^{2}/Q^{2}$, where $Q^{2}$
is a typical squared Euclidean four-momentum of $\Gamma$,
and such an expansion would never provide an adequate representation of
$\Gamma$ in the region $Q^{2} \le \Lambda^{2}$, whereas the dynamical
equations determining $\Gamma'$s contain loop integrals whose evaluation
requires a systematic approximation of those $\Gamma'$s over the entire
momentum range. Of necessity, the index $r$ will therefore have to refer
to a sequence of more {\it{global}} approximants, whereas for $g^{2}$,
which enters the dynamical equations only parametrically, a local
approximation around $g^{2} = 0$ is possible. The peculiar asymmetry
between the two "directions" of the sequence is therefore not an
arbitrary choice, but is rooted in the very nature of the spontaneous
scale (1.2).
 
On the other side,
the meaning of $p$ will also be subtly different from what it is in the
purely perturbative context, since for $p \ge 1$ it will
refer to corrections calculated from diagrammatic building blocks
$\Gamma^{[r,0]}$, rather than from the standard Feynman-rules vertices
$\Gamma^{(0) {\it{pert}}}$ implied in eq. (1.1). In other words,
as compared to perturbation theory, the approximation will be able to
"correct its own zeroth order". Determination of the
set $\{ \Gamma^{[r,0]}\}$, the {\it{nonperturbatively modified vertices
of zeroth perturbative order}}, will then lead to the self-consistency
problem outlined in sect. 4, which is one of the characteristic new
features of the scheme. We discuss the remarkable fact that this
self-consistency problem becomes {\it{rigorously}} restricted, not by
any "decoupling" approximations but by the very nature of a mechanism
tied to the divergence structure of the theory, to a small finite set
of vertices, and we indicate how perturbative renormalization
continues to function in the framework of this mechanism.
Finally, section 5 has a number of comments on the special description
emerging in this context of the elementary QCD excitations. Apart from a
few glimpses provided by the sample calculation of sect. 4, this outline
does not dwell at all on either the bulk of the actual loop and
self-consistency computations, nor on the interesting, but nevertheless
dynamically secondary, question of approximate nonperturbative
saturation of
the Slavnov-Taylor identities, since both subjects are of an algebraic
lengthiness that requires separate presentation. We do, however, comment
occasionally on the relation with the work of refs. [7], which the
present scheme allows to put in perspective as a lowest stage (although
with somewhat arbitrary technical simplifications) of a more systematic
sequential approximation.
 
\subsection{Notation and Conventions}
The following notation will be used. One considers correlation functions
generated by the standard gauge-theory action
in $D = 4 - 2 \epsilon$ Euclidean dimensions,
\begin{equation}
S = \int \, d^{D} x \left[ {\cal{L}}_{V} (x) + \sum \limits_{F}
{\cal{L}}_{F} (x) + {\cal{L}}_{G} (x) \right]
\end{equation}
where ${\cal{L}}_{V}$ and ${\cal{L}}_{F}$ are Lagrange densities,
respectively, of a set of $SU(N_{C})$ gauge-vector (gluon) fields and
of a set of minimally coupled fermion (quark) fields coming in $N_{F}$
flavors $F = u, d, s \ldots$, while the term
\begin{equation}
{\cal{L}}_{G} = \frac{1}{2 \xi_{0}} \left[ \partial^{\mu} A^{\mu}_{a}
(x) \right]^{2} + {\bar{c}}_{a} (x) \left\{
\left[-\delta_{ab} \partial^{\mu} + {\tilde{g}}_{0} f_{abc} A^{\mu}_{c}
(x) \right] \partial^{\mu} \right\} c_{b} (x) \, .
\end{equation}
comprises standard covariant gauge-fixing and Fadde'ev-Popov (FP) terms.
The coupling factor ${\tilde{g}}_{0} = \nu^{\epsilon}_{0}
g_{0}$, with $\nu_{0}$ another mass scale, is used to keep the
bare gauge coupling $g_{0}$ dimensionless. We will focus in particular
on the connected, amputated and one-particle irreducible correlation
functions ({\it{proper vertices}}) with $N$ external lines in Euclidean
momentum space,
\begin{equation}
\Gamma_{N} (\{k\}), \qquad \qquad \{k\} = \{k_{1}, k_{2}, \ldots
k_{N} | k_{1} + k_{2} + \ldots + k_{N} = 0 \} \, .
\end{equation}
Where necessary, the $N$ external lines will be labelled more
specifically by $V$ for a vector (gluon) line -- occasionally detailed
further as $T$ or $L$ for a transverse or longitudinal gluon,
respectively --, by $G, {\bar{G}}$ for ghost and antighost lines, and by
$F, {\bar{F}}$ for quark and antiquark lines. Thus $\Gamma_{TLV}$ will
be a purely gluonic three-point vertex with one transverse, one
longitudinal, and one generic gluon line, $\Gamma_{FVV{\bar{F}}}$ a
quark-gluon-gluon-antiquark four-point vertex, etc. The numbers of the
three types of lines will be written $n_{V}, n_{G}, n_{F}$, so that
\begin{equation}
n_{V} + n_{G} + \sum \limits_{F} \, n_{F} = N \, .
\end{equation}
In particular, the set of proper two-point vertices
\begin{equation}
\Gamma_{2} : = \left\{ \, \, \Gamma_{VV} = -D^{-1},
\, \, \, \, \Gamma_{G{\bar{G}}} =
-{\tilde{D}}^{-1}, \, \, \, \, \left\{ \Gamma_{F{\bar{F}}} =
-S^{-1}_{F}
| F = 1 \ldots N_{F}\right\} \, \, \right\} \, ,
\end{equation}
consists of the negative inverses of the vector propagator $D$, ghost
propagator
$\tilde{D}$, and fermion propagators $S_{F}$, while the set
\begin{equation}
\Gamma_{3} : = \left\{ \, \, \Gamma_{3V}, \, \, \Gamma_{GV{\bar{G}}},
\, \, \, \, \left\{
\Gamma_{FV{\bar{F}}} | F = 1 \ldots N_{F} \right\} \, \, \right\}
\end{equation}
comprises the basic three-point interaction vertices. The dynamics
implied by the action (1.7 -- 1.8) are then embodied in the
Dyson-Schwinger (DS) equations [8], an infinite, hierarchical system of
coupled integral equations for the functions (1.9) (see [9] for their
specific form in QCD), which in a condensed notation take the form
\begin{equation}
\Gamma_{N} = \Gamma^{(0) {\rm{pert}}}_{N} + \left(\frac{g_{0}}{4 \pi}
\right)^{2} \Phi_{N} \left[ \Gamma_{2}, \Gamma_{3}, \ldots, \Gamma_{N},
\Gamma_{N + 1}, \Gamma_{N + 2} \right] \, .
\end{equation}
The {\it{perturbative zeroth-order or bare vertices}}, $\Gamma^{(0)
{\rm{pert}}}_{N}$, are given by the standard Feynman rules for the
action (1.7 - 1.8), while $\Phi_{N}$ denotes a set of nonlinear dressing
functionals, defined by loop integrals over combinations of $\Gamma's$,
and the notation emphasizes that each such loop integral is preceded
by at least one power of the bare gauge coupling, $g^{2}_{0}$.
 
It is important to keep in mind that a fundamental dichotomy is
introduced into the set of vertices (1.9) through the renormalizable
divergence structure of a QCD-like theory. A small finite subset,
consisting of the {\it{superficially divergent or basic vertices}},
\begin{equation}
\Gamma_{s div} : = \left\{ \, \, \Gamma_{2}, \, \, \Gamma_{3}, \,
\, \Gamma_{4V} \, \,
\right\}\, ,
\end{equation}
are distinguished by the fact that loop integrals in their $\Phi_{N}$
functionals of eqs. (1.13) have a non-negative value of the overall
degree of divergence in $D = 4 \, \, [10]$,
\begin{equation}
\delta_{N} = 4 - n_{V} - \frac{3}{2} \left(n_{G} + \sum \limits_{F}
n_{F}\right) \, ,
\end{equation}
so that each of these need their own specific renormalizations, which
can be performed at least perturbatively [11]. Related to this is the
fact that it is precisely for $\Gamma_{sdiv}$ that the bare terms
$\Gamma^{(0) {\rm{pert}}}$ in eq. (1.13) are nonzero. By contrast, the
remaining, infinite set of {\it{superficially convergent or higher
vertices}},
\begin{equation}
\Gamma_{s conv} : = \left\{ \, \, \Gamma_{GVV{\bar{G}}},
\, \, \Gamma_{FVV{\bar{F}}}, \, \, \Gamma_{FF{\bar{F}}{\bar{F}}}, \, \,
\Gamma_{5V},\, \,
\ldots
\right\}
\end{equation}
have loop integrals with a negative $\delta_{N}$, and therefore exhibit
no typical divergences of their own, but at most subdivergences
representing corrections to the basic vertices (1.14): when rewritten
as {\it{dressed-skeleton expansions}}, i.e. in terms of fully dressed
and renormalized basic vertices (1.14), their loops are actually
convergent.
 
While testable consequences of the theory are mostly contained
in (the color-singlet channels of) the higher Green's functions (1.16),
these cannot be calculated in a truly systematic way without first
studying, and renormalizing, the basic vertices (1.14): a
disproportionate amount of theoretical effort must be directed towards a
class of amplitudes that contain next to nothing in observable physics.
(An exception may be the four-gluon vertex $\Gamma_{4V}$ in (1.14),
which
may develop glueball poles.) The approach described below is, at its
present stage, concerned exclusively with the DS equations (1.13) for
the seven superficially divergent vertices (1.14).
 
The usual {\it{perturbative solution}} to eqs. (1.13) is obtained by
straightforward iteration around $\Gamma_{N}^{(0) {\rm{pert}}}$ and by
applying, at each step, a renormalization scheme $R$, which among other
things eliminates $g_{0}$ in favor of a renormalized coupling
$g(\nu)$ depending on the arbitrary scale $\nu$ :
\begin{equation}
\Gamma^{\rm{pert}}_{N} = \lim \limits_{p \rightarrow \infty} \, \,
\Gamma^{[p] {\rm{pert}}}_{N}; \quad \quad
\Gamma_{N}^{[p] \rm{pert}} = \Gamma^{(0) {\rm{pert}}}_{N} +
\sum \limits^{p}_{p' = 1} \, \left[ \frac{g (\nu)}{4 \pi}
\right] ^{2p'} \Gamma^{(p') {\rm{pert}}}_{N} \, .
\end{equation}
Here the radiative corrections $\Gamma^{(p) {\rm{pert}}}_{N} \, \,
(p \ge 1)$ are computed iteratively from the zeroth-order solution
$\Gamma^{(0) {\rm{pert}}}_{N}$, i. e. from the standard Feynman rules.
For example, the first iteration is described schematically by
\begin{equation}
\left\{ \left( \frac{g_{0}}{4 \pi} \right)^{2} \Phi_{N} \left[
\Gamma^{(0) {\rm{pert}}} \right] \right\}_{R,\nu} =
\left[ \frac{g (\nu)}{4 \pi} \right]^{2} \Gamma^{(1) {\rm{pert}}}_{N}
+ 0 (g^{4}) \, .
\end{equation}
For $R$, we always have in mind a
dimensional-regularization-plus-minimal-subtraction scheme with respect
to $D = 4 - 2 \epsilon$, which in particular entails the familiar
coupling renormalization
\begin{equation}
g^{2}_{0} \nu^{2 \epsilon}_{0} = g^{2} (\nu) \nu^{2 \epsilon} Z_{\alpha}
(g^{2}(\nu), \epsilon) \, ,
\end{equation}
where at the one-loop level
\begin{equation}
Z_{\alpha} = 1 - \beta_{0} \left[\frac{g(\nu)}{4 \pi} \right]^{2}
\frac{1}{\epsilon} + 0 (g^{4}) \, ; \qquad
\beta_{0} = \frac{11}{3} N_{c} - \frac{2}{3} N_{F} \, .
\end{equation}
 
\subsection{Problems not Addressed}
It may be clarifying to mention two related problems which, while
important in their own right, are {\it{not}} addressed in this paper.
 
{\underline{(i)}} It is by no means clear at this time whether (1.7)
with (1.8) is the correct action to use in continuum QCD.
Recent sharpening by Zwanziger [12,13] of Gribov's criticism [14] of the
insufficient gauge fixing afforded by (1.8) has rendered untenable the
convenient prejudice that "Gribov's problem may be ignored in
perturbative treatments". It is now quite clear that if one insists
on quantizing the chromodynamic system with a non-redundant set of
degrees of freedom for the gauge field, the ensuing restriction of the
path integral to a fundamental domain, containing exactly one
representative of each gauge orbit, affects also the "perturbative"
region of small fields. Zwanziger has in fact presented [12] a method
of approximately "exponentiating" the fundamental-domain restriction,
which demonstrates the minimum of new effects to be expected from a more
complete gauge fixing. His treatment leads to the replacement of the
action (1.7) by $S' = S + \gamma H$, where $\gamma$ is a parameter of
dimension mass$^{4}$ and of the non-analytic type (1.3), while $H$ is a
nonlocal "horizon" functional,
\begin{equation}
H = \int \, d^{D} x d^{D} y A^{\mu}_{a} (x) f_{cae} \left[
M^{-1} (x,y) \right]_{cd} f_{dbe} A^{\mu}_{b} (y) \, ,
\end{equation}
with $M^{-1}$ denoting the inverse of the FP operator in the curly
bracket of (1.8). (This nonlocal term may be replaced with a local one
by using path integration over additional auxiliary fields.) The drastic
low-energy effects produced by this term -- in particular, the emergence
of the gluon propagator of eq. (2.17) below, describing a short-lived
elementary excitation of the gluon field -- bear several intriguing
similarities to those emerging from the "extended perturbative" approach
discussed here, but the precise relationship of the two methods is
unknown. What should be emphasized at this time is that there
is not necessarily a contradiction between them. The Dyson-Schwinger
solution
for the "redundantly" described system of eqs. (1.7/1.8), when given
enough nonperturbative freedom, may settle down self-consistently in
the region singled out by the non-redundant description, and the recent
insight that the effect of (1.21) can be described in terms of a
nonperturbative vacuum spontaneously breaking BRS invariance [15]
may be seen as pointing in this direction.
 
{\underline{(ii)}} The present paper studies a purely Euclidean theory.
In the few instances where we need to refer to its properties in the
Minkowskian domain, as in sect. 5 below, we shall proceed, like most
current investigations of QCD, on the assumption that the usual strategy
of Euclidean field theory -- to define Green's functions and solve for
them entirely in the Euclidean, and to continue to the Minkowskian only
in the final answers -- yields physically correct results.
For the gauge-fixing dependent correlations of the {\it{elementary}}
QCD fields, from which any analytic investigation has to start,
there seems to be at present no full proof of this, and the discussion
below will show that in the approximating subsequence of primary
interest for QCD, the Euclidean solution continued
analytically will differ from the corresponding direct solution to the
Minkowskian equations. Since the core parts of the
method described here, and in particular the basic quantum effect
leading to self-consistency of the nonperturbative terms, are general
enough to continue to work in e.g. a purely
Minkowskian theory, we have not given this question any priority, but
its existence should be kept in mind when proceeding to various
applications that the method will invite.
 
\section{The Extended Perturbative Expansion}
\subsection{General Restrictions}
\setcounter{equation}{0}
 
We next describe the form and general properties of the extended
iterative sequence, with no attention as yet to the question of how
this form achieves self-consistency in the DS equations of the theory.
In this sequence, as the terminology indicates, the organizing principle
of the perturbative solution (1.17) is not discarded altogether: one
still considers a formal power series in the parameter $[g (\nu)/4
\pi]^{2}$, and therefore a weak-coupling solution that is directly
applicable (i.e., applicable without infinite resummations) only if that
parameter remains sufficiently small {\it{at all scales $\nu$}}
to permit semi-convergent expansions of the $\Gamma_{N}$. It is worth
emphasizing that this assumption is entirely  compatible with present
fragmentary knowledge about the flow of $g$ in QCD: what is truly known
of $g(\nu)$ are a few leading orders of its asymptotic expansion
at large $\nu$, which are of the inverse-logarithmic form of eq.
(1.4), with $k^{2}$ replaced by $\nu^{2}$.
As for low-$\nu$ behavior, although folklore about "the
running coupling blowing up at the scale $\Lambda$" has become so
pervasive as to be occasionally confused with theory,
the emerging consensus, if any, from phenomenology [16] and lattice
studies [17] seems rather to point in the direction anticipated by
Gribov
in 1987 [18]: as $\nu \rightarrow 0$, the running coupling does nothing
dramatic, but "freezes" around an order of magnitude of perhaps
0.2 for the quantity $[g(\nu)/2 \pi]^{2} = \alpha_{s}/\pi$ in the
$\overline{\rm{MS}}$ scheme.
In the present context,
as we shall see, {\it{amplitudes}} can get large at momenta $\approx
\Lambda$, but via a rather different route opened up by the novel
feature of the expansion. The new feature is that
we now have a double, two-index sequence of approximants, in which each
term is allowed an additional dependence on the $g^{2}$-nonanalytic
mass parameter (1.3):
\begin{equation}
\Gamma_{N} (\{k\}; g^{2} (\nu); \nu) = \lim \limits_{r \rightarrow
\infty} \, \, \lim \limits_{p \rightarrow \infty} \, \Gamma^{[r,p]}_{N}
( \{k\}; g^{2} (\nu), \nu) \, ,
\end{equation}
\begin{equation}
\Gamma^{[r,p]}_{N} = \Gamma^{[r,0]}_{N} (\{k\}; \Lambda) + \sum
\limits^{p}_{p'=1} \, \left[ \frac{g(\nu)}{4 \pi} \right]^{2p'}
\Gamma^{\left[\left. r,p' \right)\right.}_{N} (\{k\}; \Lambda; \nu)
\end{equation}
(We are using an abbreviated notation suppressing all dependence on
parameters not immediately relevant to the present argument, such
as quark masses and gauge-fixing parameters.) One may immediately state
two {\it{boundary conditions}} on the nonperturbatively modified
$\Gamma^{[r,p]}$ amplitudes. Since $\Lambda^{2}$, by (1.3), vanishes
faster as $g^{2} \rightarrow 0$ than any power of $g^{2}$, it makes
physical sense to consider the (formal) limit in which $\Lambda^{2}
\rightarrow 0$ but the $g^{2p}$ remain finite. In this {\it{perturbative
limit}} we should ensure
\begin{equation}
\Gamma^{\left[\left. r,p \right) \right.} (\Lambda^{2} = 0) =
\Gamma^{(p) {\rm{pert}}} \qquad (p = 0,1,2 \ldots) \, .
\end{equation}
On the other hand, since QCD is asymptotically free, and since the
logarithmic corrections to asymptotic freedom are known to arise
from partial resummation of the $g^{2}$-power series, it is plausible
to demand that the zeroth-order functions $(p=0)$ should possess
{\it{naive}} asymptotic freedom, i.e.
\begin{equation}
\Gamma^{[r,0]} (\{\lambda k\}) \rightarrow \Gamma^{(0) {\rm{pert}}}
(\{\lambda k\}), \qquad \lambda \rightarrow \infty \, ,
\end{equation}
as the set $\{k\}$ of external four-momenta are scaled up uniformly.
These simple-looking conditions will be seen to strongly restrict the
nonperturbative extension. We will actually impose a restriction
slightly stronger than (2.4), namely the requirement that
\begin{equation}
\begin{array}{cr}
the \, \, \, nonperturbative \, \, \, extension \, \, \,
of \, \, \, \Gamma^{\rm{(pert)}} & \mbox{}\\
should \,\, \, continue \, \, \,
to \, \, \, be \, \, \, perturbatively \, \, \, renormalizable. &
\mbox{}
\end{array}
\end{equation}
In an asymptotically free theory, and only there, this requirement
is intuitively plausible: the large-momentum behavior of vertex
functions in the loop integrals of (1.13) is known to be essentially
the perturbative one, apart from slowly varying logarithmic
modifications that should not change the divergence pattern
qualitatively. What makes (2.5) a somewhat stronger statement (for
vertices with $N \ge 3$) is the implied
condition that behavior no worse than for the perturbative vertex should
obtain even when {\it{only those momenta of $\Gamma$ that run in a loop
become large}}, while the remaining, "external" ones are kept
constant.
Comparison between eqs. (3.8) and (3.9) below, for the three-gluon
vertex, will pin down this difference more quantitatively in a specific
example.
 
We already emphasized that in the "nonperturbative direction" of the
sequence (2.1), characterized by the index $r$ and relating to the
dependence on $\Lambda^{2}$, the approximation, in
contrast to the local one provided by a Taylor series around a point,
must be {\it{global}}. On the other hand, at the basis of any
perturbative renormalization process lies the possibility of
{\it{superficial convergence assessment by integer-power counting}}.
The only simple meeting point for these two requirements, and for the
direction provided by the examples in sect. 1 of nonperturbative
quantities, is the idea of representing $\Gamma^{[r,0]}$, the
nonperturbatively modified vertex functions of zeroth perturbative
order, by {\it{rational approximants in $\Lambda^{2}$}} of increasing
order $r$. This will indeed be seen to lead to a viable approximation
scheme.
 
\subsection{Gluonic Two-Point Function}
In this section, we give details for the simplest case of a scalar
vertex function with only one invariant momentum argument: the two-point
vertex, or negative-inverse propagator, of transverse gluons. In the
notation explained in sect. 1, this is
\begin{equation}
\Gamma_{T} (k^{2}) = - \frac{1}{D_{T} (k^{2})} \, ,
\end{equation}
where $D_{T}$ is the invariant function defined by the tensor
decomposition of the full Euclidean gluon propagator,
\begin{equation}
D^{\mu \nu} (k) = t^{\mu \nu} (k) D_{T} (k^{2}) + l^{\mu \nu} (k)
D_{L}\left(k^{2}\right) \, ;
\end{equation}
\begin{equation}
t^{\mu \nu} (k) = \delta^{\mu \nu} - \frac{k^{\mu} k^{\nu}}
{k^{2}} = \delta^{\mu \nu} - l^{\mu \nu} (k) \, ;
\end{equation}
\begin{equation}
D_{L}(k^{2}) = \frac{\xi_{0}}{k^{2}}\, .
\end{equation}
The propagation characteristics of the gluonic elementary excitation,
$A^{\mu}_{a} (x) |0>$, will be determined by the zeroes and branch
points of $\Gamma_{T}$ in the complex $k^{2}$ plane. In the Euclidean
domain, its rational approximants, which we will
characterize by their {\it{denominator}} degrees $r$, will be of the
form
\begin{equation}
\begin{tabular}{r@{\, \, = \, \,}l}
\large{
$-\Gamma^{[r,0]}_{T} (k^{2}, \Lambda^{2})$} & \Large{
{$\frac{(k^{2})^{r+1} + \zeta_{r,1} \Lambda^{2} (k^{2})^{r} +
\ldots + \zeta_{r, r+1} (\Lambda^{2})^{r+1}}
{(k^{2})^{r} + \eta_{r,1} \Lambda^{2} (k^{2})^{r-1} + \ldots
+ \eta_{r,r} (\Lambda^{2})^{r}}\, ; $}}\\ [4ex]
$r$ & $0,1,2,3, \ldots$ \, ;
\end{tabular}
\end{equation}
where the $2 r + 1$ coefficients $\zeta_{r,i}$ and $\eta_{r,j}$ are all
real.
Note how the boundary condition (2.4), in which
\begin{equation}
-\Gamma^{(0) {\rm{pert}}}_{T} (k^{2}) = k^{2} \, ,
\end{equation}
has uniquely fixed both the relative degrees and the leading
coefficients of the numerator and denominator, and how condition (2.3)
is then automatically fulfilled. {\it{Only in an asymptotically free
theory does one have such strong a priori restrictions on the form of the
approximants.}}
 
Without imposing specific dynamics, the sequence (2.10) still covers a
variety of physical situations. We do not give a complete classification
here, which would include several unphysical or exotic cases, but
mention just the two subsequences of primary physical interest:
 
(1) "Particle" subsequence. Here $r$ is even, so that the propagation
function
$D_{T}^{[r,0]} (k^{2})$, by (2.6), has odd denominator degree $r +
1$, and therefore at least one pole on the real $k^{2}$ axis, since the
coefficients are real. (Note our convention of using the
rational-approximation index $r$ also on the propagator, although for
the latter it gives the numerator, rather than the denominator,
degree.) If the real pole closest to the origin sits
at timelike Minkowskian (i.e. negative Euclidean) $k^{2}$, it represents
a stable, asymptotically detectable gluon particle. Assuming -- as
everybody seems to have assumed tacitly since the classic papers of
Lehmann and K\"allen [19] -- that the elementary operator field can
connect the vacuum to at most one single-particle state, one would
expect this mass-shell pole position to stabilize as $r$ is increased.
The remaining $r$ poles and $r$ numerator zeroes of $D^{[r,0]}_{T}$
would then be expected to come again on the real axis, but separated
from the particle pole, and for increasing $r$ would be expected to
settle into the alternating pattern that in the context of rational
approximants is known [20] to approximate a branch cut -- the
Lehmann-K\"allen dressing cut, arising from virtual decays of the
particle into multiparticle configurations (Fig. 1A).
 
The simplest approximant of this sequence, $r = 0$, is
\begin{equation}
-\Gamma^{[0,0]}_{T} (k^{2}, \Lambda^{2}) = k^{2} + \zeta_{0,1}
\Lambda^{2} \qquad (\zeta_{0,1} \, \, {\rm{real \, \, and}} \, > 0)
\end{equation}
which when compared to (2.11) represents the general Schwinger mechanism
[21]: the spontaneous creation of a mass $m^{2} = \zeta_{0,1}
\Lambda^{2}$ of the nonperturbative type (1.2) in a massless bare
propagator.
 
(2) "Quasiparticle" subsequence. Here $r$ is odd, so that
$D^{[r,0]}_{T}$
has an even number $r + 1$ of poles, and at least one real zero.
Ignoring again the exotic possibility of two or more stable-gluon poles
at different real masses, one would expect
the two poles closest to the origin to come as a complex-conjugate
pair at, say, $k^{2} = - \sigma_{r, \pm} \Lambda^{2},$
with $\sigma_{r,+} = \sigma_{r,-}^{*}$
being a dimensionless complex number. As discussed in [22] and in
sect. 5 below, the conceptual problems apparently caused by this
complex-poles structure are not insurmountable, provided the solution is
used consistently. The pair would represent an
{\it{intrinsically
short-lived elementary excitation}} of the gauge-vector field, with
lifetime of the order of $1/\Lambda$. Note that the real zero closest
to the origin -- corresponding to a pole of the function (2.6), i.e., a
singular gluonic self-energy -- is every bit as essential in this
context as the complex pole pair, as it will be seen to make the
connected Green's functions nonsingular in the invariant masses
of external vector lines. The remaining even number $r - 1$ both
of poles and of zeroes may then come in complex-conjugate pairs,
farther from the origin than the leading "quasiparticle" pair, with
poles and zeroes interspersed so as to approximate two conjugate
branch lines (Fig. 1B). These again would represent dressing --
the short-lived excitation coupling dynamically to multiple copies of
itself. The absence of real-axis branchpoints would signal that it has
no channels for decay into stable fragments that would jointly
carry the open quantum numbers of a gluon. This
structure would fit most closely
the empirical situation for the short-lived gluon presumed to be present
at the origin of a gluon-jet event.
 
The simplest approximant of this sequence, $r = 1$, is
\begin{eqnarray}
 - \Gamma^{[1,0]}_{T} (k^{2}, \Lambda^{2}) & = &
\frac{(k^{2})^{2} + \zeta_{1,1} \Lambda^{2} k^{2} + \zeta_{1,2}
\Lambda^{4}}{k^{2} + \eta_{1,1} \Lambda^{2}} \nonumber \\
& = & k^{2} + u_{1,1} \Lambda^{2} + \frac{u_{1,3}
\Lambda^{4}}{k^{2} + u_{1,2} \Lambda^{2}}
\end{eqnarray}
with real coefficients $u_{1,i}$ given by
\begin{displaymath}
u_{1,1} = \zeta_{1,1} - \eta_{1,1} ; \qquad
u_{1,2} = \eta_{1,1} ; \qquad
u_{1,3} = \zeta_{1,2} - \eta_{1,1} u_{1,1} ;
\end{displaymath}
and satisfying
\begin{equation}
u_{1,3} > \left[ \frac{1}{2} (u_{1,1} - u_{1,2}) \right]^{2} \, .
\end{equation}
The corresponding nonperturbatively modified propagator of zeroth
perturbative order,
\begin{equation}
D^{[1,0]}_{T} (k^{2}, \Lambda^{2}) =
\frac{k^{2} + u_{1,2} \Lambda^{2}}{(k^{2} + \sigma_{1,1} \Lambda^{2})
(k^{2} + \sigma_{1,3} \Lambda^{2})} \, ,
\end{equation}
exhibits the minimum of features mentioned above: a real zero at
$k^{2} = - u_{1,2} \Lambda^{2}$, and a "quasiparticle" pair of
complex-conjugate poles at $k^{2} = -\sigma_{1,\pm} \Lambda^{2}$, where
\begin{equation}
\sigma_{1,+} = \sigma_{1,1} = \frac{1}{2} \left( u_{1,1} + u_{1,2}\right)
+ i \sqrt{u_{1,3} - \left[ \frac{1}{2} (u_{1,1} - u_{1,2}) \right]^{2}}
= \sigma^{*}_{1,3} = \sigma^{*}_{1,-}
\end{equation}
At $r = 1$, there are no further
zeroes and poles as yet to simulate dressing cuts.
 
Eqs. (2.15) and (2.13) are of a form suggested in [22] (for $u_{1,2} =
0$) and used in [7] as an element of an approximate, nonperturbative DS
solution, and that form can now be identified as the lowest member of
(the odd-$r$ subsequence of) a systematic sequence (2.10) of
approximants to the nonperturbatively modified function (2.6). A special
case,
\begin{equation}
D^{(GZ)}_{T} (k^{2}) = \frac{k^{2}}{k^{4} + \gamma} \, ,
\end{equation}
with $u_{1,1} = u_{1,2} = 0$ and $u_{1,3}  \Lambda^{4} = \gamma$,
had actually been arrived at much earlier by Gribov [14], and was later
derived independently by Zwanziger [12], via the entirely different
route of fundamental-domain restrictions. Because of this different
origin, the Gribov-Zwanziger $\gamma$ term is present {\it{already
at tree level}}, i.e. in the analog of what we called $\Gamma^{(0)
{\rm{pert}}}$, whereas the mechanism to be used in sect. 4 to stabilize
nonzero $u_{1,i}$ coefficients in (2.13) will be seen to operate from DS
{\it{loops}}.
 
It is clearly desirable to try out all of the above types of
approximants in the DS equations, to determine those that can achieve
dynamical self-consistency and, if necessary, to further distinguish
between the latter by a stability analysis. We do not here embark
on such a comprehensive study, but describe a few steps toward the much
more restricted program of trying out the "quasiparticle" subsequence
-- with odd $r$ and {\it{only}} complex-conjugate propagator
singularities. In this subsequence, the nonperturbatively modified
two-vector vertex in zeroth perturbative order will be of the form
\begin{eqnarray}
- \Gamma^{[r,0]}_{T} (k^{2}, \Lambda^{2}) & = & k^{2} + u_{r,1}
\Lambda^{2} + \frac{u_{r,3} \Lambda^{4}}{k^{2} + u_{r,2}
\Lambda^{2}} \nonumber \\
 & & + \sum \limits^{(r-1)/2}_{s=1} \left[ \frac{u_{r,4s+1}
\Lambda^{4}}{k^{2} + u_{r, 4s} \Lambda^{2}} + \frac{u_{r, 4s+3}
\Lambda^{4}}{k^{2} + u_{r, 4s+2} \Lambda^{2}} \right]  \\
\bigskip \nonumber  \\
 r & = & 1,3,5 \ldots
\end{eqnarray}
generalizing eq. (2.13), with $u_{r,1}, u_{r,2}, u_{r,3}$ real, with
\begin{equation}
u_{r, 4s+2} = u^{*}_{r, 4s}, \qquad
u_{r, 4s+3} = u^{*}_{r, 4s+1} \qquad
\left(s = 1 \ldots \frac{r-1}{2} \right) \, ,
\end{equation}
and such that all poles of the corresponding propagation function,
\begin{equation}
D^{[r,0]}_{T} (k^{2}) = \frac{k^{2} + u_{r,2} \Lambda^{2}}
{(k^{2} + \sigma_{r,1} \Lambda^{2}) (k^{2} + \sigma_{r,3} \Lambda^{2})}
\prod \limits^{(r-1)/2}_{s=1}
\frac{(k^{2} + u_{r, 4s} \Lambda^{2}) (k^{2} + u_{r, 4s+2} \Lambda^{2})}
{(k^{2} + \sigma_{r, 4s+1} \Lambda^{2}) (k^{2} + \sigma_{r, 4s+3}
\Lambda^{2})}
\, ,
\end{equation}
form complex-conjugate pairs, with $\sigma_{r,1}, \sigma_{r,3}$ denoting
the leading
"quasiparticle" pair closest to the origin of the $k^{2}$ plane,
i.e.,
\begin{equation}
\sigma_{r,3} = \sigma^{*}_{r,1};\, \,  \sigma_{r, 4s+3} = \sigma^{*}_{r,
4s+1};
\quad
|\sigma_{r, 4s+1}| > |\sigma_{r,1}| \quad (s = 1 \ldots \frac{r-1}{2})
\, .
\end{equation}
The naive-asymptotic-freedom condition built into (2.10),
$\Gamma^{[r,0]}_{T} \rightarrow -k^{2}$ for $k^{2} >\!> \Lambda^{2}$,
guarantees that in the pole decomposition,
\begin{equation}
D^{[r,0]}_{T} (k^{2}) = \frac{\rho_{r,0}}{k^{2} + \sigma_{r,1}
\Lambda^{2}}
+ \frac{\rho_{r,2}}{k^{2} + \sigma_{r,3} \Lambda^{2}} + \sum
\limits^{(r-1)/2}_{s=1} \left[ \frac{\rho_{r, 4s}}{k^{2} + \sigma_{r,
4s+1}
\Lambda^{2}} + \frac{\rho_{r, 4s+2}}{k^{2} + \sigma_{r, 4s+3}
\Lambda^{2}}
\right] \, ,
\end{equation}
the sum of the dimensionless residues is unity:
\begin{equation}
\rho_{r,0} + \rho_{r,2} + \sum \limits^{(r-1)/2}_{s=1} (\rho_{r,4s}
+ \rho_{r, 4s+2}) = 1 \quad (r = 1,3,5 \ldots) \, .
\end{equation}
This, incidentally, is one of the features that distinguish the
functions (2.23) from propagators
plagued by so-called ghost poles, which -- usually as a result of
inadequate approximations -- vexed field theorists in the 1950's [23],
and for which the sum-of-residues was negative or vanishing.
 
\subsection{Relation with the OPE}
To exhibit the connections of the sequential approximation (2.10) with
an established area of QCD methodology, we briefly look at the
operator-product expansion (OPE). The OPE for the
{\it{elementary}} QCD fields has only recently begun to be
established correctly (for recent results and literature, see refs.
[24] through [27]) and is not commonly discussed in the present terms,
yet it already represents a step towards describing the
"truly nonperturbative" $\Lambda^{2}$ dependence. Again we consider the
simplest case
of the one-variable vertex function (2.6), and restrict ourselves to a
theory with at most massless quarks, so as not to have to worry about
the role of other invariant mass scales besides $\Lambda$. By writing
an OPE for the Euclidean two-point function of the gauge field,
transforming to momentum space, contracting with the $t^{\mu \nu}$ of
eq. (2.8) to project out the transverse portion, and forming the inverse
function (2.6), one arrives at an expansion of the general form
\begin{equation}
\begin{tabular}{rl}
$-\Gamma_{T} (k^{2}; g^{2} (\nu); \nu) \quad = $ & \quad
$k^{2} \left\{ 1 + \sum \limits^{\infty}_{p=1} \left[\frac{g (\nu)}
{4 \pi} \right]^{2p} L_{0,p} \left( \frac{k^{2}}{\nu^{2}} \right)
\right\}$\\ [4ex]
\mbox{} &
$+ \Lambda^{2} \left\{ L_{10} + \sum \limits^{\infty}_{p=1}
\left[\frac{g (\nu)}
{4 \pi} \right]^{2p} L_{1,p} \left( \frac{k^{2}}{\nu^{2}} \right)
\right\}$\\ [4ex]
\mbox{} &
$+ \frac{\Lambda^{4}}{k^{2}} \left\{ L_{20} + \sum
\limits^{\infty}_{p=1} \left[\frac{g (\nu)}
{4 \pi} \right]^{2p} L_{2,p} \left( \frac{k^{2}}{\nu^{2}} \right)
\right \}$\\ [4ex]
\mbox{} & $ + \ldots$\\ [4ex]
\mbox{} & $+ \frac{(\Lambda^{2})^{n}}{(k^{2})^{n-1}} {\Biggl\{} L_{n,0}
+ \ldots {\Biggr\}}$\\[4ex]
\mbox{} & $+ \ldots$
\end{tabular}
\end{equation}
where the coefficient functions (modified Wilson coefficients) are
$p-th$ degree polynomials of logarithms:
\begin{equation}
L_{n,p} = c^{(0)}_{n,p} + c^{(1)}_{n,p} \ln
\left(\frac{k^{2}}{\nu^{2}} \right) + \ldots +
c^{(p)}_{n,p} \left[ \ln \left( \frac{k^{2}}{\nu^{2}} \right)
\right]^{p} \, .
\end{equation}
This may be viewed as a form of the series (2.1/2.2) where the
$\Gamma^{\left[\left. r,p \right)\right.}$ have in turn been expanded in
power series in $\Lambda^{2}$. More precisely, apart from the typical
perturbative logarithms, it is an expansion (presumably
semi-convergent) in powers of $(\Lambda^{2}/k^{2})$ for
$k^{2} >\!> \Lambda^{2}$, the deep-Euclidean limit. The first line of
(2.25) is identical with the perturbation series (1.17), so the boundary
condition (2.3) is satisfied. The presence of the other terms,
containing powers of the $g^{2}$-nonanalytic scale (1.2), shows clearly
that the perturbative series alone would be an incomplete solution even
when summed to all orders. At the core of these additional terms are
the quantities $L_{n0} (\Lambda^{2})^{n}, n \ge 1$, which the OPE
derivation identifies as linear combinations of vacuum expectation
values of composite operators ("vacuum condensates") of increasing
mass dimension, $2n$.
 
As it stands, the OPE (2.25) does not satisfy our needs, for two
reasons.
First, the OPE by itself does not determine the quantities $L_{n0}$ --
this requires a truly dynamical principle, such as the DS equations.
Second, even if the $L_{n0}$ were determined dynamically up to some
$n$, no finite order of (2.25) would be a satisfactory continuation of
the vertex function into the region of primary physical interest --
the region $k^{2} \le \Lambda^{2}$ of typical hadronic masses. In this
region, given our assumption that $[g (\nu)/4\pi]^{2}$ never becomes
large, the important task clearly is to obtain a
continuation-through-resummation of the "vertical" sums in (2.25) --
in particular, the $p=0$ vertical sum,
\begin{equation}
- \Gamma^{[r,0]}_{T} (k^{2}, \Lambda^{2}) = k^{2} \left[
1 + L_{10} \frac{\Lambda^{2}}{k^{2}} + L_{20} \left( \frac{\Lambda^{2}}
{k^{2}} \right)^{2} + \ldots \right] \, ,
\end{equation}
which is {\it{free of perturbative logarithms.}}
(Once this is done,
the remaining $(p \ge 1)$ "vertical" summations
can in principle be generated
iteratively from $\Gamma^{[r,0]}$ through the DS equations, just as
in the perturbative case.) From this standpoint, the rational
approximations (2.10) may now be viewed as a systematic sequence of
continuations-through-resummation of the OPE subseries (2.27), which
locate its low-$k^{2}$ zeroes and singularities with increasing
accuracy.
 
We stress that the OPE (2.25) has been
used here only as a point of comparison.  The
"extended-iterative"
scheme differs from it in more than the technical aspect of
parametrization. (In the OPE, the fundamental parameters are an infinite
set of dimensionful vacuum condensates, whereas in (2.10)
that role is played by the dimensionless vertex coefficients such as
$(\zeta_{r,i}, \eta_{r,j})$, and condensates are secondary quantities
calculable in principle in terms of the latter -- see the second of
refs. [7] for $[r,p] = [1,0]$ examples). Its properties can be
qualitatively different from any finite order of (2.25) because the two
are separated by a nontrivial step of analytic continuation. In
particular, the "horizontal" sums in (2.25) are always ordinary QCD
perturbation series based on the Feynman rules, with all the
attendant problems arising from
the empirically wrong zeroth-order spectrum (including free, massless,
physical gluons, and their associated infrared singularities) of the
latter. By contrast, expansion (2.1/2.2) is based on the use of (inter
alia) modified propagators such as (2.21), which in particular produce
no infrared singularities at all (the unphysical longitudinal gluons
still do, but their effects are always preceded by $\xi$ factors
that identify them as gauge-fixing artefacts). This opens up the
possibility that Borel transforms with respect to $g^{2}$ of eq. (2.2)
may have, apart from gauge-fixing artefacts, no genuine infrared
renormalons, the effects of the latter having been absorbed in a
redefinition of the zeroth perturbative order.
 
\section{Three-Vector-Vertex Approximants}
\setcounter{equation}{0}
 
The hierarchical structure of the DS equations (1.13) implies that in
principle all proper vertices $\Gamma_{N}$ should be treated
simultaneously by mutually consistent, nonperturbative approximants.
The method described here will however be found to have the
simplifying feature that the essential self-consistency problem -- for
the vertices $\Gamma^{[r,0]}$ -- is {\it{rigorously}} restricted to the
small finite set (1.14) of superficially divergent vertices. In this
section we therefore compile formulas for rational functions
$\Gamma^{[r,0]}_{sdiv}$, concentrating on the
example of the proper three-vector vertex $\Gamma_{3V}$, which serves to
illustrate all the essential features. The largely analogous material
for the remaining $\Gamma_{sdiv}$ vertices will be relegated to appendix
A. The only new aspect this discussion will turn up will be the
"factorizing-denominator" rule discussed in connection with eq. (3.6)
below. Otherwise, the main complications will be the multi-variable
nature, and the notoriously unwieldy color-and-Lorentz-tensor structure,
of the $N \ge 3$ QCD vertices.
One measure we will adopt to control this purely
kinematical complexity is to restrict ourselves to vertices with at
most transverse external gluon lines, wich are sufficient for performing
calculations in the Landau $(\xi_{0} = 0)$ gauge. As shown most clearly
by the example of eq. (2.9), it is in these that nonperturbative effects
are expected to develop most freely. Amplitudes with at least one
longitudinal gluon are strongly restricted by the Slavnov-Taylor (ST)
identities, whose nonperturbative saturation
is not a subject of this paper.
 
The three-gluon vertex has color structure,
\begin{equation}
\left(\Gamma_{3V}\right)^{\mu \lambda \nu}_{abc} = f_{abc}
\Gamma^{\mu
\lambda
\nu}_{3V (f)} + d_{abc} \Gamma^{\mu \lambda \nu}_{3V (d)} \, ,
\end{equation}
in terms of f-type (antisymmetric) and d-type (symmetric) structure
constants. Each color component in turn is of the general Lorentz
structure discussed by Ball and Chiu [28]: a linear combination of
14 independent third-rank tensors, with 6 different invariant functions
having appropriate symmetry or antisymmetry properties in the
Lorentz-scalar variables $p^{2}_{1}, p^{2}_{2}, p^{2}_{3}$. We consider
only the totally transverse portions,
\begin{equation}
\Gamma^{\mu' \lambda' \nu'}_{3T(c)} = t^{\mu' \mu} (p_{1}) t^{\lambda'
\lambda} (p_{2}) t^{\nu' \nu} (p_{3}) \Gamma^{\mu \lambda \nu}_{3V (c)}
\qquad (c = f \, \, {\rm{or}} \, \, d) \, ,
\end{equation}
which have contributions from only 4 Lorentz tensors combined with two
different invariant functions $F_{0}, F_{1}$ for each color component:
\begin{displaymath}
\Gamma^{\mu' \lambda' \nu'}_{3T (c)} (p_{1}, p_{2}, p_{3}) =
t^{\mu' \mu} (p_{1}) t^{\lambda' \lambda} (p_{2}) t^{\nu' \nu}
(p_{3}) \times
\end{displaymath}
\begin{displaymath}
\times \left\{ \delta^{\lambda \nu} (p_{2} - p_{3})^{\mu}
F_{(c) 0} (p^{2}_{2}, p^{2}_{3}, p^{2}_{1}) \right.
\end{displaymath}
\begin{equation}
+ \delta^{\mu \nu} (p_{3} - p_{1})^{\lambda} F_{(c) 0}
(p^{2}_{3}, p^{2}_{1}, p^{2}_{2})
\end{equation}
\begin{displaymath}
+ \delta^{\mu \lambda} (p_{1} - p_{2})^{\nu} F_{(c) 0} (p^{2}_{1},
p^{2}_{2}, p^{2}_{3})
\end{displaymath}
\begin{displaymath}
\left. + (p_{2} - p_{3})^{\mu} (p_{3} - p_{1})^{\lambda} (p_{1} -
p_{2})^{\nu}
F_{(c) 1} (p^{2}_{1}, p^{2}_{2}, p^{2}_{3}) \right\}; \qquad \,\, (c = f
\,\,
{\rm{or}} \, \, d) \, .
\end{displaymath}
Remember $p_{1} + p_{2} + p_{3} = 0$. Here the dimensionless functions
$F_{(c) 0}$ are symmetric (for $c = f$) or antisymmetric (for $c = d$)
in their first two arguments, while the functions $F_{(c) 1}$, of mass
dimension $-2$, are totally symmetric ($c = f$) or antisymmetric
($c = d$) in all three arguments. The perturbative zeroth-order limits
are,
\begin{equation}
F^{(o) {\rm{pert}}}_{(c) k} = \delta_{cf} \delta_{k0} \qquad
(c = f \, \, {\rm{or}} \, \, d, \quad k = 0 \, \, {\rm{or}} \, \, 1) \,
.
\end{equation}
In setting up sequences of rational approximants for these four
invariant functions, capable of dynamical consistency with the
gluon-propagator sequence (2.18), one encounters a new aspect:
approximants with the most general denominator polynomials in the
three variables $p^{2}_{1}, p^{2}_{2}, p^{2}_{3}$ are not useful.
The zeroes of such a general denominator in any one variable $p^{2}_{i}$
are complicated non-rational functions of the two other variables, and
this will stand in the way of DS self-consistency when a vertex
transfers its structure in an external momentum to the next lower
vertex through the hierarchical coupling. At the expense of slower
convergence of the approximating sequences, we must restrict ourselves
to the narrower but still sufficiently general class of
{\it{factorizing-denominator rational approximants (FDRA)}}, i.e. those
in which the denominator factorizes with constant zeroes in all three
variables. To see that such a more special approximation is always
possible in principle, consider e.g. the function
$F_{(f) 0} (p^{2}_{1}, p^{2}_{2}; p^{2}_{3})$ as the kernel of a
symmetric integral operator parametrically dependent on $p^{2}_{3}$,
and write an eigenfunction expansion
\begin{equation}
F_{(f) 0} (p^{2}_{1}, p^{2}_{2}; p^{2}_{3}) = \sum \limits_{n}
g_{n} (p^{2}_{3}) f_{n} (p^{2}_{1}) f_{n} (p^{2}_{2}) \, ,
\end{equation}
with eigenvalues $g_{n}$. By using single-variable rational
approximation for the $f_{n}$ and $g_{n}$ functions, letting  $n$
range over finite but increasing numbers of eigenvalues, suitably
discretizing the integral in case there is a continuous spectrum,
and putting everything over a common denominator, one generates a
sequence of approximants in which denominators have the desired, fully
factorized form. Similar considerations apply to the other invariant
functions.
 
By extension of our above definition, the degree $r$ of rational
approximation for the three-point vertex $\Gamma^{[r,0]}_{3T}$ in
zeroth perturbative order
will be the number of different denominator zeroes in any one variable
{\it{for the entire tensorial vertex}}. That is, to order $p = 0$ in
eq. (2.1) and degree $r$, all invariant functions will be of the form
\begin{equation}
F^{[r,0]}_{(c) k} (p^{2}_{1}, p^{2}_{2}, p^{2}_{3}) =
\frac{N^{(r)}_{3V (c) k} (p^{2}_{1}, p^{2}_{2}, p^{2}_{3})}
{\left[ \prod \limits^{r}_{s=1} \left(p^{2}_{1} + u'_{r, 2s}
\Lambda^{2}\right) \right] \left[ \prod \limits^{r}_{s=1} \left(
p^{2}_{2} + u'_{r, 2s} \Lambda^{2} \right) \right]
\left[ \prod \limits^{r}_{s=1} \left(p^{2}_{3} + u'_{r, 2s}
\Lambda^{2} \right) \right]}
\end{equation}
\begin{displaymath}
(c = f \, \, {\rm{or}} \, \, d; \quad k = 0 \, \, {\rm{or}} \, \, 1;
\quad r = 1, 3, 5, \ldots )
\end{displaymath}
with the same fully factorized denominator, where the appropriate
mass dimensions and Bose -- symmetry restrictions, as well as
the boundary conditions (2.3/2.4), are built into the numerator
polynomials
\begin{equation}
N^{(r)}_{3V (c) k} \left(p^{2}_{1}, p^{2}_{2}, p^{2}_{3}\right) =
\delta_{cf}
\delta_{k0} \left(p^{2}_{1} p^{2}_{2} p^{2}_{3}\right)^{r}
+ \sum \limits_{m_{1}, m_{2}, m_{3} \ge 0} \, C^{(c)k}_{m_{1} m_{2}
m_{3}}
\left( p^{2}_{1} \right)^{m_{1}}
\left( p^{2}_{2} \right)^{m_{2}} \left( p^{2}_{3} \right)^{m_{3}}
\left( \Lambda^{2} \right)^{3r - k - (m_{1} + m_{2} + m_{3})} \, .
\end{equation}
Condition (2.4) allows nonzero coefficients $C_{m_{1} m_{2} m_{3}}$
only for
\begin{equation}
m_{1} + m_{2} + m_{3} \le 3r - k - 1 \, .
\end{equation}
Additional restrictions follow from the postulate (2.5) of preservation
of perturbative renormalizability. As a minimum, it requires that
no part of $\Gamma_{3V}$ should lead to ultraviolet divergences stronger
than the corresponding perturbative ones. Since in a $1PI$ diagram one
of the three legs of $\Gamma_{3V}$ may be external, we conclude that
when any two of its momenta are running in a loop, the vertex
should behave no worse than $q^{1}$ at large loop momenta $q$. For the
invariant functions (3.6) this requires
\begin{equation}
m_{1} + m_{2} \le 2r - k, \quad
m_{2} + m_{3} \le 2r - k, \quad
m_{3} + m_{1} \le 2r - k \, .
\end{equation}
Moreover, the coefficients should exhibit the symmetries with respect
to $m_{1}, m_{2}, m_{3}$ necessary to fulfill the conditions of partial
or total (anti-)symmetry listed before.
 
Comparison between eqs. (3.9) and (3.8) illustrates well the statements
made after eq. (2.5): of the conditions required for a general rational
function to reduce to one that preserves perturbative power counting,
the vast majority are already enforced by the asymptotic-freedom
condition (3.8), with (3.9) representing only a mild additional
restriction. (In fact, on the $r = 1$ level written out below, (3.9)
will already be fully implied by (3.8).)
For the discussion of sect. 4, the essential and simple consequence of
the factorizing-denominator structure is that with respect to any one
argument $p^{2}_{k}$, the full $p = 0$ vertex has a partial-fraction
decomposition
\begin{displaymath}
\left[ \Gamma^{\mu_{1} \mu_{2} \mu_{3}}_{3T (c)} (p_{1}, p_{2}, p_{3})
\right]^{[r,0]} = B^{\mu_{i} \mu_{j} \mu_{k}}_{(c), 0, r} (p_{i}, p_{j},
p_{k})
\end{displaymath}
\begin{equation}
+ \sum \limits^{r}_{s=1} B^{\mu_{i} \mu_{j} \mu_{k}}_{(c) s, r}
(p_{i}, p_{j}) \left( \frac{\Lambda^{2}}{p^{2}_{k} + u'_{r,2s}
\Lambda^{2}} \right)
\end{equation}
\begin{displaymath}
(i,j,k = 1,2,3 \, \, {\rm{and \, \, cyclic)}} \, ,
\end{displaymath}
where invariant functions for the residue tensors $B_{s,r}$ with
$s \ge 1$ depend only on $p^{2}_{i}$ and $p^{2}_{j}$, whereas the
regular part $B_{0,r}$ is in addition an $r$-th order polynomial in
$p^{2}_{k}$. (Note that with respect to any {\it{single}} $p^{2}_{k}$,
the conditions (3.8/3.9) do not rule out terms that grow polynomially.
Without dwelling on this point, we remark that most of these terms will
however turn out to vanish on dynamical grounds.)
 
The question may be raised as to whether the choice of variables
$p^{2}_{1}, p^{2}_{2}, p^{2}_{3}$ in constructing the FDRA of eq. (3.6)
is unique. Could other sets of Lorentz invariants be used, such
as $\{ p_{1} \cdot p_{2}, p_{2} \cdot p_{3}, p_{3} \cdot p_{1} \}$?
In eq. (3.6), we have, for brevity, anticipated the fact that dynamical
consistency between the $\Gamma_{2V}$ and $\Gamma_{3V}$ vertices will
indeed require that FDRA's for $\Gamma_{3V}$ be constructed in the
$p^{2}_{k}$ variables.
 
To illustrate these structures, we write expressions (3.6), now
fully decomposed into partial fractions in analogy with eq. (2.18),
for the simplest case, $r = 1$. In this case, all rational structure can
be expressed in terms of the single-variable pole factors
\begin{equation}
\Pi_{i} = \frac{\Lambda^{2}}{p^{2}_{i} + u'_{1,2} \Lambda^{2}} \qquad
(i = 1,2,3)\,.
\end{equation}
The f-type invariant-function approximants then read,
\begin{equation}
\begin{array}{rl}
F^{[1,0]}_{(f)0} \left(p^{2}_{1}, p^{2}_{2}; p^{2}_{3}\right) = &
1 + x_{1,1} \left(\Pi_{1} + \Pi_{2} \right) + x_{1,3} \Pi_{3}
+ \left(x_{1,2} + x'_{1,2} \frac{1}{\Pi_{3}} \right) \Pi_{1}
\Pi_{2}\\[4ex]
\mbox{} & + \left[ \left( x_{1,4} + x'_{1,4} \frac{1}{\Pi_{2}} \right)
\Pi_{1} + \left( x_{1,4} + x'_{1,4} \frac{1}{\Pi_{1}} \right)
\Pi_{2} \right] \Pi_{3}\\[4ex]
\mbox{} & + x_{1,5} \left( \Pi_{1} \Pi_{2} \Pi_{3} \right) \, ,
\end{array}
\end{equation}
\begin{equation}
F^{[1,0]}_{(f) 1} \left(p^{2}_{1}, p^{2}_{2}, p^{2}_{3} \right) =
\frac{1}{\Lambda^{2}} \left[x_{1,6} \left( \Pi_{1} \Pi_{2} +
\Pi_{2} \Pi_{3} + \Pi_{3} \Pi_{1} \right) + x_{1,7}
\left( \Pi_{1} \Pi_{2} \Pi_{3} \right) \right] \, ,
\end{equation}
whereas the d-type approximants are,
\begin{equation}
F^{[1,0]}_{(d)0} \left(p^{2}_{1}, p^{2}_{2}; p^{2}_{3} \right) =
x_{1,8} \left( \Pi_{1} - \Pi_{2} \right) + \left[ \left( x_{1,9}
+ x'_{1,9} \frac{1}{\Pi_{2}} \right) \Pi_{1} - \left(x_{1,9} +
x'_{1,9}
\frac{1}{\Pi_{1}} \right) \Pi_{2} \right] \Pi_{3} \, ,
\end{equation}
\begin{equation}
F^{[1,0]}_{(d)1} \left(p^{2}_{1}, p^{2}_{2}, p^{2}_{3} \right)
\equiv 0 \, .
\end{equation}
The latter result reflects the fact that for $k = 1$ and $r = 1$, no
fully antisymmetric numerator polynomial in three variables obeying
restriction (3.8) can be constructed.
 
To order $(r = 1, p = 0)$, the set of twelve dimensionless coefficients,
\begin{equation}
x_{1} = \{ x_{1,1}, x_{1,2}, \ldots x'_{1,9} \} \, ,
\end{equation}
which must be real numbers of order unity, are the fundamental vertex
parameters to be determined from the DS equations (1.13). Note that all
invariant functions at this level feature one single pole position,
$-u'_{1,2} \Lambda^{2}$. While e.g. in the $k = 0$ functions the
partial (anti-) symmetry would seem to permit the use of two different
pole positions, such an approximant, under our above definition, would
already be of degree $r=2$. This strict definition of $r$ may at first
appear to be overly rigid, but it will turn out to be the relevant one
for the self-consistency problem.
 
\section{Dyson-Schwinger self-consistency}
\setcounter{equation}{0}
 
When introduced into the DS equations, the nonperturbative
sequence (2.1), in contrast to the purely perturbative one, encounters
a nontrivial self-consistency problem -- the self-reproduction of the
$p = 0$ functions $\Gamma^{[r,0]}$ at a given level $r$. For the first
iteration, and in the notation of eq. (1.13), this problem can be
stated as
\begin{equation}
\left\{ \left( \frac{g_{0}}{4 \pi} \right)^{2} \Phi_{N} \left[
\Gamma^{[r,0]} \right] \right\}_{R, \nu} = \left[ \Gamma^{\left[ \left.
r,0 \right) \right.}_{N} - \Gamma^{(0){\rm{pert}}}_{N} \right] +
0 \left(g^{2} (\nu), e (r + 1) \right) \, .
\end{equation}
Here, $e (r + 1)$ stands for the approximation error in the $r$
direction which, in contrast to that of the semiconvergent $p$
sequence, cannot be characterized by a power of some expansion
parameter.
 
It is useful to reflect at this point on what the precise meaning can
be of
"solving a Dyson-Schwinger system by a sequence $\Gamma^{[r,p]}$":
 
(1) Upon iteration of the system around the starting point
$\Gamma^{[r,0]}$, the sequence should reproduce itself at each iteration
step {\it{up to corrections of the next higher perturbative order $p$}},
which are simultaneously generated in the process.
 
(2) Self-reproduction of $\Gamma^{[r,0]}_{N}$, which features a number
$n(r)$ of dimensionless nonperturbative parameters, can only mean that
this approximant is made to coincide with the r.h.s. of its own DS
equation at $n(r)$ points in the space of scalar momentum variables
$k^{2}$ of $\Gamma_{N}$, or more generally (since global approximants
allow a variety of matching prescriptions) {\it{with respect to $n(r)$
"comparison data"}} (function values, derivatives, pole positions,
integrated quantities $\ldots$). At all other points, or in all other
data, a matching error will necessarily remain that can be improved only
by going to the next higher $r$.
 
\subsection{Transverse-Gluon Self-Energy}
It is characteristic of the scheme discussed here that the hardest
problem for direct numerical Dyson-Schwinger solutions -- the
self-reproduction of {\it{momentum}} structure -- has a relatively
straightforward and explicit answer: the $(N' > N)$-point vertices
entering the dressing functional $\Phi$ will "hand down" their rational
structure in external momenta to the N-point ones. This pattern may at
first sight appear simplistic, but in fact turns out to be an efficient
way of exploiting the most difficult and peculiar structural feature
of a Dyson-Schwinger system -- the hierarchical coupling. It has,
however, the immediate consequence that the set of nonperturbatively
modified vertices can only be treated {\it{as a whole}}: no
self-consistency, however approximate, will be possible in this
framework if one seeks solutions to particular DS equations while
treating the higher vertices appearing there by more or less
unrelated assumptions.
 
Less straightforward is self-reproduction of the {\it{coupling}}
structure. This problem is already clearly visible in (4.1): the DS loop
integrals constituting $\Phi_{N}$ must produce the bracket on the
r.h.s., with no $g^{2}$ prefactor, in spite of the fact that they
always come with at least one $g^{2}_{0}$ prefactor. The importance of
securing this feature can hardly be overstated: since perturbative
corrections of finite order $(p \ge 1)$ cannot alter the qualitative
spectral properties of a solution, the essential nonperturbative
features, and in particular qualitative changes expected in the
spectrum of elementary excitations -- i.e., in the two-point
functions of the basic fields -- must establish themselves already on
the $p = 0$ level.
 
The mathematical mechanism leading to the "eating of $g^{2}_{0}$
prefactors", a low-order version of which was used already in [22]
and discussed in detail in [7], will again be demonstrated for the
example of the one-variable function (2.6). Its DS equation,
\begin{equation}
\Gamma_{T} (k^{2}) = - k^{2} + \left( \frac{g_{0}}{4 \pi} \right)^{2}
\sum \limits_{M = A \ldots F} \, \, \Phi^{(M)}_{TT}
\left[ \Gamma_{2}, \Gamma_{3}, \Gamma_{VVVT} \right] \, ,
\end{equation}
is stated diagrammatically in fig. 2, showing the one-DS-loop
$(M = A, B, C, D)$ and two-DS-loop terms ($M = E,F$) of the $\Phi_{TT}$
functional. (The term "DS loop" will be used to refer to the specific
structure of DS integrals which are neither bare nor fully dressed,
containing always one bare vertex times a number of dressed functions).
Start at the one-loop level, where the terms $M = E,F$ do not yet
contribute. For eq. (4.1), we have to evaluate
\begin{equation}
\begin{array}{rl}
\Phi^{(l = 1)}_{TT} \left[ \Gamma^{[r,0]} \right] = &
\Phi^{(A)}_{TT} \left[ D^{[r,0]}, \Gamma^{[r,0]}_{VTV} \right] +
\Phi^{(B)}_{TT} \left[ {\tilde{D}}^{[r,0]}, \Gamma^{[r,0]}_{GT
\overline{G}} \right]\\[4ex]
\mbox{} & + \Phi^{(C)}_{TT} \left[ D^{[r,0]} \right] +
\sum \limits_{F} \, \, \Phi^{(D)}_{TT} \left[ S^{[r,0]}_{F}, \,
\Gamma^{[r,0]}_{FT \overline{F}} \right] \, .
\end{array}
\end{equation}
Now use the FDRA structure of the $[r,0]$ three-point vertices, as
expressed in the partial-fraction decompositions (3.10), (A.24), and
(A.44). Expression (4.3) then decomposes into terms regular and singular
with respect to the external variable $k^{2}$:
\begin{equation}
\Phi^{(l=1)}_{TT} \left[ \Gamma^{[r,0]} \right] = I^{(r)}_{0} (k^{2}) +
\sum \limits^{r}_{s=1} \, I^{(r)}_{s} (k^{2}) \left(
\frac{\Lambda^{2}}{k^{2} + u'_{r, 2s} \Lambda^{2}} \right) \, .
\end{equation}
The quadratically divergent loop integrals $I^{(r)}_{s}$ are of the
same form as (4.3), but with the replacements
\begin{equation}
\Gamma^{[r,0]}_{VTV} \longrightarrow B^{(r)}_{(f)s} \quad {\rm{in}}
\quad \Phi^{(A)}_{TT} \, ,
\end{equation}
\begin{equation}
\Gamma^{[r,0]}_{GT\overline{G}} \longrightarrow {\tilde{B}}^{(r)}_{(f)s}
\quad
{\rm{in}}
\quad \Phi^{(B)}_{TT} \, ,
\end{equation}
\begin{equation}
\Gamma^{[r,0]}_{FT{\overline{F}}} \longrightarrow C^{(r)}_{(F)s} \quad
{\rm{in}}
\quad \Phi^{(D)}_{TT} \, .
\end{equation}
Indices $(f)$ on the r.h. sides of (4.5/4.6) recall the fact that only
the $f_{abc}$ parts of those vertices contribute here. By definition,
the "tadpole" contribution $\Phi^{(C)}_{TT}$,
(which does not vanish as in the perturbative case),
is entirely included in $I^{(r)}_{0}$, as it has
no $k^{2}$-singular terms, and in fact is a constant.
 
In eq. (4.4), since we have treated
the 3-point vertices consistently, i.e. by FDRA's of the same level
$r$, the r.h.s. already displays the needed number of poles, and
inspection of the $I^{(r)}_{s}$ integrals shows that they produce
logarithmic branch points (generally complex since we are using
the complex-pole propagators of the odd-r sequence) but no poles
in $k^{2}$ of their own. It is then natural to choose the $r$
positions and $r$ residues of those poles as comparison data -- if the
poles were not matched, the matching error would be locally infinite.
That is, one requires
\begin{equation}
u_{r, 2s} = u'_{r,2s} \quad (s = 1 \ldots r) \, ;
\end{equation}
\begin{equation}
- \, u_{r, 2s + 1} \Lambda^{2} = \left( \frac{g_{0}}{4 \pi} \right)^{2}
I^{(r)}_{s} \left( -u_{r,2s} \Lambda^{2} \right) \quad (s = 1 \ldots r)
\, .
\end{equation}
Eq. (4.8) expresses the "handing down" of momentum structure, rationally
approximated, from the 3-point vertices to the 2-point one under
consideration. It immediately implies, by eq. (2.21), that the
propagator will have {\it{zeroes}} at the positions of the
gluon-variable poles of the 3-point vertices.
 
It would seem that as a $(2r + 1)$-th comparison datum to fix the
one still undetermined coefficient $u_{r,1}$ of (2.18), we could use
the value of the smooth-remainder function,
\begin{equation}
J^{(r)}_{0} (k^{2}) = I^{(r)}_{0} (k^{2}) + \Lambda^{2} \, \sum
\limits^{r}_{s=1}
\,
\frac{I^{(r)}_{s} (k^{2}) - I^{(r)}_{s} (-u_{r,2s}
\Lambda^{2})}{k^{2} + u_{r,2s} \Lambda^{2}} \, ,
\end{equation}
at some arbitrary point, $k^{2} = -{\overline{u}} \Lambda^{2}$, of
the $k^{2}$ plane:
\begin{equation}
- \, u_{r,1} \Lambda^{2} = \left( \frac{g_{0}}{4 \pi} \right)^{2}
J_{0}^{(r)} (- {\overline{u}} \Lambda^{2}) \, .
\end{equation}
We will soon see that the arbitrariness introduced at this point is only
apparent. In eqs. (4.9) and (4.11), the problem of how to "eat" the
$g^{2}_{0}$ prefactor to produce a $p = 0$ quantity is still with us.
 
\subsection{The self-consistency mechanism}
Evaluating now the $I_{s}$ and $J_{0}$ integrals in dimensional
regularization -- note that all these integrals, because of the rational
structure of integrands, can be evaluated by standard methods --, one
finds
\begin{equation}
I^{(r)}_{s} (-u_{r,2s} \Lambda^{2}) = \left(
\frac{\Lambda^{2}_{\epsilon}}{\nu^{2}_{0}} \right)^{-\epsilon} \left\{
a_{s} (u, v, w, x, y, z) \cdot \frac{1}{\epsilon} + [{\rm{terms \, \,
finite \, \, as \, \,}} \epsilon \rightarrow 0] + 0 (\epsilon) \right\}
\Lambda^{2} \quad (s = 1 \ldots r) \, ,
\end{equation}
\begin{equation}
\hspace{-0.5cm} J^{(r)}_{0} (-{\overline{u}} \Lambda^{2}) = \left(
\frac{\Lambda^{2}_{\epsilon}}{\nu^{2}_{0}} \right)^{-\epsilon} \left\{
a_{0} (u, v, w, x, y, z) \cdot \frac{1}{\epsilon} + [{\rm{terms \, \,
finite \, \, as \, \,}} \epsilon \rightarrow 0] + 0 (\epsilon) \right\}
\Lambda^{2}  \, ,
\end{equation}
where the notation recalls that the $\Lambda$ scale, too, should be
continued to $D = 4 - 2 \epsilon$ through the replacement $\beta(g')
\rightarrow \beta(g') -
\epsilon g'$ in its definition (1.2). The sets $u \ldots
z$ are the dimensionless-coefficient sets of the $[r,0]$ propagators
and 3-point vertices, as defined in sect. 3 and appendix A, and include,
for the time being, the quantity $\overline{u}$ of eq. (4.11).
 
To analyze the product $(g_{0}/4 \pi)^{2}
(\Lambda^{2}_{\epsilon}/\nu^{2}_{0})^{- \epsilon} (1/\epsilon)$ now
appearing in the self-consistency equations, we next impose the
condition, inherent in (2.5), that perturbative coupling-constant
renormalization, eq. (1.19), should apply. Then this product becomes,
\begin{equation}
\left[ \frac{g (\nu)}{4 \pi} \right]^{2} \, Z_{\alpha} \left( g^{2}
(\nu), \frac{1}{\epsilon} \right) \left( \frac{\Lambda^{2}_{\epsilon}}
{\nu^{2}} \right)^{-\epsilon} \frac{1}{\epsilon} \equiv \Pi (\epsilon,
g^{2}) \, .
\end{equation}
To get a feeling for its structure, we first use a slightly sloppy
argument [7,6] which, however, happens to convey the essential points.
Start from
\begin{equation}
\left( \frac{\Lambda^{2}_{\epsilon}}{\nu^{2}}\right)^{-\epsilon} =
1 - \epsilon \ln \left( \frac{\Lambda^{2}}{\nu^{2}} \right) +
O (\epsilon^{2}) \, ,
\end{equation}
which gives
\begin{equation}
\frac{1}{\epsilon} \left( \frac{\Lambda^{2}_{\epsilon}}{\nu^{2}}
\right)^{-\epsilon} = - \ln \left( \frac{\Lambda^{2}}{\nu^{2}} \right)
+ \frac{1}{\epsilon} + O (\epsilon) \, .
\end{equation}
The logarithmic term survives the removal-of-regulator limit, $\epsilon
\rightarrow 0$,
in the same, familiar way as do the perturbative logarithms of, say, eq.
(2.25), but it differs from them in one important respect -- it leads
them by one order in the perturbative $g^{2p}$ classification. From eq.
(1.3), at the one-loop level
\begin{equation}
-\ln \left( \frac{\Lambda^{2}}{\nu^{2}} \right) = \left[ \frac{4 \pi}
{g(\nu)} \right]^{2} \frac{1}{\beta_{0}} \left\{ 1 + O (g^{2} \ln
g^{2}, g^{4}) \right\} \, .
\end{equation}
This is precisely the $1/g^{2}$ factor needed to "eat" the overall
$g^{2}$ factor in (4.14); it has survived due to its association with
the $1/\epsilon$ divergence. We now have
\begin{equation}
\left[ \frac{g (\nu)}{4 \pi} \right]^{2} \frac{1}{\epsilon} \left(
\frac{\Lambda^{2}{\epsilon}}{\nu^{2}} \right)^{-\epsilon} =
\frac{1}{\beta_{0}} \left\{ 1 + \left[ \frac{g (\nu)}{4 \pi} \right]^{2}
\beta_{0} \frac{1}{\epsilon}
+ O (g^{2} \ln g^{2}, g^{4}) + O (g^{2} \epsilon) \right\} \, .
\end{equation}
But $Z_{\alpha}$, to the same order, is given by (1.20), so that to the
order calculated, the divergences actually cancel in the product (4.14):
\begin{equation}
Z_{\alpha} \left\{ \left[ \frac{g (\nu)}{4 \pi} \right]^{2}
\frac{1}{\epsilon} \left( \frac{\Lambda^{2}_{\epsilon}}{\nu^{2}}
\right)^{- \epsilon} \right\} = \frac{1}{\beta_{0}} \left[ 1 + O (g^{2}
\ln g^{2}) + O (g^{4}) + O (\epsilon) \right]
\end{equation}
Although the terms denoted $O (g^{4})$ do include ( as do those in
(1.20) ) terms of the form $g^{4}/\epsilon$, they are not to be kept
in a one-loop calculation. We therefore find that not only do the
nonperturbative parts of (2.18) -- the $r$ pole terms and the mass-type
constant term -- establish themselves in order $p = 0$ in spite of the
$p = 1$ prefactor, but they do so without divergences to the order
calculated, so that in particular {\it{no nonlocal counterterms are
needed}} for the pole terms. This crucial prerequisite of perturbative
renormalization is therefore automatically preserved.
 
The sloppiness of the above derivation based on 1-loop results lies
hidden in the terms summarily denoted "$O(\epsilon^{2})$" in (4.15)
and "$O(g^{4})$" in (1.20). When examining these more closely, one
realizes that $(\Lambda^{2}_{\epsilon}/\nu^{2})^{-\epsilon}$ actually
has an infinite subseries of terms of type $(\epsilon/g^{2})^{m},
m \ge 1$, whereas $Z_{\alpha}$ has a subseries of terms of type
$(g^{2}/\epsilon)^{n}, n \ge 0$, and the desired terms of type
$\epsilon^{0} (g^{2})^{0}$ in (4.14) can therefore be produced in an
infinite
number of ways. To account for all of these, one could obtain both
subseries exactly by resummations of the $LL$ type, but the result,
while perfectly sufficient for the one-loop calculation, would not
display the basic simplicity of the situation: the quantity $\Pi$ of
(4.14) is actually independent of $g^{2}$. This follows immediately from
the exact integral representations of the two main factors. From the
extension of definition (1.2) to $\epsilon \not= 0$, one finds by simple
manipulations that
\begin{equation}
\left(\frac{\Lambda^{2}_{\epsilon}}{\nu^{2}} \right)^{-\epsilon} =
\frac{\kappa_{1}}{\kappa (\nu)} \exp \left\{ \int
\limits^{\kappa(\nu)}_{\kappa_{1}} \frac{d \lambda}{\lambda + \epsilon
f(\lambda)} \right\} \, ,
\end{equation}
where $\kappa = \left[ g(\nu)/4 \pi \right]^{2}$, and
\begin{equation}
f(\kappa) = \left[ \beta_{0} + \beta_{1} \kappa + \beta_{2} \kappa^{2} +
\ldots \right]^{-1} = \frac{- g \kappa}{\beta (g)}\, .
\end{equation}
The $1/\kappa$ factor of (4.20) is the exact counterpart of the one of
(4.17) and is what is needed for the "eating" process. The
lower integration limit $\kappa_{1}$, fixed conventionally as part
of the definition of the renormalization scheme, will in general depend
on $\epsilon$,
\begin{equation}
\kappa_{1} = \kappa_{1} (\epsilon) = \kappa_{1} (0) [1 + O(\epsilon)]
\, ,
\end{equation}
with $\kappa_{1} (0)$ having the meaning of a renormalized coupling at
the scale $\Lambda$. On the other hand, differentiating (1.19) with
respect to $\nu$ leads to the relation
\begin{equation}
\kappa \frac{d}{d \kappa} \ln Z_{\alpha} (\kappa (\nu), \epsilon) =
- \frac{1}{1 + \frac{\epsilon}{\kappa} f (\kappa)} \, ,
\end{equation}
whose reintegration {\it{with respect to $\kappa$}} under
the initial condition $Z_{\alpha} (0, \epsilon) = 1$ gives
\begin{equation}
Z_{\alpha} (\kappa (\nu), \epsilon) = \exp \left\{ - \int
\limits^{\kappa(\nu)}_{0} \frac{d \lambda}{\lambda + \epsilon f
(\lambda)}
\right\} \, ,
\end{equation}
a representation known to t'Hooft [29] in 1973. Now the dependence
on $\kappa (\nu)$
cancels exactly in the product of (4.20) and (4.24); the result
\begin{equation}
\Pi (\epsilon) = \frac{\kappa_{1}}{\epsilon} \exp \left\{ - \int
\limits^{\kappa_{1}(\epsilon)}_{0} \frac{d \lambda}{\lambda + \epsilon f
(\lambda)} \right\}
\end{equation}
has none of the
apparent $g^{2}$ corrections of (4.19). That $\Pi(\epsilon)$ is,
moreover, finite as $\epsilon \rightarrow 0$  -- the exact
counterpart of the divergence cancellation observed on the way
from (4.18) to (4.19) -- follows by writing
\begin{equation}
\frac{1}{\lambda + \epsilon f (\lambda)} = \frac{1}{\lambda +
(\epsilon/\beta_{0})} + \epsilon \varrho_{1} (\lambda, \epsilon) \, ,
\end{equation}
where $\varrho_{1} (0, \epsilon) = 0$; then
\begin{equation}
\Pi (\epsilon) = \frac{1}{\beta_{0}} \left[ 1 +
\frac{\epsilon}{\beta_{0} \kappa_{1} (\epsilon)} \right]^{-1}
\exp \left\{ - \epsilon \int^{\kappa_{1} (\epsilon)}_{0} d \lambda
\varrho_{1} (\lambda, \epsilon) \right\} \, .
\end{equation}
Further subtraction of $\varrho_{1}$ at $\lambda = 0$ shows that the
integral develops also a term $\ln \epsilon$, so finally,
\begin{equation}
\Pi (\epsilon) = \frac{1}{\beta_{0}} \left[ 1 + O (\epsilon, \epsilon
\ln \epsilon) \right] \, .
\end{equation}
This is a compact formulation of the self-consistency mechanism. It
clearly shows, through the $\frac{1}{\epsilon}$ factor in (4.14), that
the mechanism is {\it{tied to the divergence structure}} of the theory,
and represents a dynamical exploitation of that structure. In this it is
reminiscent, of course, of quantum anomalies, except that here the
process seems to have no connection with the breaking of a classical
symmetry.
 
The result is that by imposing the algebraic self-consistency conditions
\begin{equation}
u_{r, 2s + 1} = - \, \frac{1}{\beta_{0}} a_{s} (u, v, w, x, y, z) \quad
(s = 0 \ldots r)
\end{equation}
in addition to (4.8), the nonperturbative terms of $\Gamma^{[r,0]}_{T}$
reproduce themselves not only without divergences but also without
finite $O (g^{2})$ corrections. Note that (4.28) contains only the
scheme-independent, leading $\beta$-function coefficient, and that it
requires evaluation only of the {\it{divergent}} parts, $a_{s}$, of
(4.12/4.13) -- the finite parts of those quantities only give terms
$O (\epsilon)$ relative to (4.29).
 
To illustrate these relations, we write conditions (4.29) for the
simplest nontrivial, odd-r case of $r=1$, in Landau gauge, and for
massless quarks, where eqs. (A.28 -- 30) apply with $\hat{m}_{F} = 0$.
They are obtained
by evaluating the one-loop quantity (4.3) with the $r = 1$ vertices
and propagators of section 3 and appendix A, using entirely standard
techniques of dimensional continuation, Feynman parametrization, and
symmetric integration, and by extracting the divergent parts as in
eqs. (4.12/4.13). One finds
\begin{eqnarray}
\beta_{0} u_{1,3} & = & \frac{N_{C}}{2} \biggl\{ 6 \left[ u_{1,1}
(x_{1,3} + 2x'_{1,4}) - x_{1,4} \right] \nonumber
\\[2ex]
\mbox{} & \mbox{} &\left.+ u_{1,2} \left[+5 x_{1,1} - \frac{5}{6}
x_{1,3} +
\frac{13}{3} x'_{1,2} + 6 x'_{1,4} - \frac{20}{3} x_{1,6} \right]
\right\}\nonumber \\[2ex]
\mbox{} & \mbox{} & + \frac{N_{C}}{2} \left\{ \frac{1}{2} (y_{1,4} +
y_{1,8})
- v_{1,1} (y_{1,2} + y_{1,5} + y_{1,9})\right.\nonumber \\[2ex]
\mbox{} & \mbox{} & + \left.\frac{1}{2} v_{1,2} (2y_{1,2} + y_{1,5} +
y_{1,9}) +
\frac{1}{6} u_{1,2} y_{1,2} \right\}\\[2ex]
\mbox{} & \mbox{} & - N_{F} \biggl\{ w_{1,3} (2 z_{1,0} - 8 z_{1,2} + 2
z_{1,3})
+ (3 w^{2}_{1,1} + 2 w_{1,1} w_{1,2} + w^{2}_{1,2})
(2 z_{1,2}
- z_{1,3})\nonumber\\[2ex]
\mbox{} & \mbox{} &+ 2 (w_{1,2} - w_{1,1}) (z_{1,1} + z_{1,2}
w_{1,2})
+ 2 z_{1,4} + (w_{1,2} + w_{1,1}) z_{1,5}
+ u_{1,2} \left[ \frac{1}{3} ( 2 z_{1,0} + z_{1,3} ) \right]
\biggr\}\nonumber\\ [5ex]
\beta_{0} u_{1,1}&  = & \frac{N_{C}}{2} \left[ -11 x_{1,1} + 6 (u_{1,1}
- x'_{1,4}) + \frac{5}{6} x_{1,3}
- \frac{13}{3} x'_{1,2} + \frac{20}{3} x_{1,6}
\right]\nonumber\\[2ex]
\mbox{} & \mbox{} & + \frac{N_{C}}{2} \left[ (v_{1,2} - v_{1,1}) +
\frac{1}{2}
(y_{1,1} + y_{1,3} - \frac{1}{3} y_{1,2}) \right]\\[2ex]
\mbox{} & \mbox{} & + \frac{N_{C}}{2} \left[ + \frac{9}{2} u_{1,1}
\right]\nonumber\\[2ex]
\mbox{} & \mbox{} &+ N_{F} \left[ - 2w_{1,3} + \frac{1}{3}
\left(2z_{1,0} + z_{1,3} \right)
+ 2 (w_{1,1} - w_{1,2}) z_{0,1}
- 2 z_{0,4} - (w_{1,1} + w_{1,2}) z_{0,5}
\right]\nonumber
\end{eqnarray}
 
plus, of course, the $u'_{1,2} = u_{1,2}$ of eq. (4.8). (An
obvious simplified notation has been employed for the
fermion-vertex coefficients $z$ of (A.45/46).) Together with
analogous conditions for the ghost and quark self-energies,
these will form an algebraic system (linear at this level) determining
$u_{1} \ldots u_{3}$,
and the other self-energy constants, in terms of the nonperturbative
coefficients $x,y,z$ of the three-point vertices.
 
After this extraction of $p = 0$ nonperturbative parts, the
remainder of the DS dressing term,
\begin{equation}
\left(\frac{g_{0}}{4 \pi} \right)^{2} \left[ J^{(r)}_{0} (k^{2}) -
J^{(r)}_{0} \left( -\overline{u} \Lambda^{2} \right) \right] \, ,
\end{equation}
still has a logarithmic divergence. In this remainder, the
quantity (4.15/20) no more occurs ( the noninteger power that
does occur contains the scale $k^{2}$ ), and treatment of its
coupling structure, under boundary conditions (2.3) and (2.5), must
therefore revert to the "normal", perturbative pattern.
Isolating the divergent piece with $R = MS$ gives,
\begin{equation}
\left( \frac{g_{0}}{4 \pi} \right)^{2} \left[ \frac{1}{2} N_{C}
\left( \frac{13}{3} \right) - \frac{2}{3} N_{f} \right]
\left(\frac{1}{\epsilon}\right)
\left( k^{2} + \overline{u} \Lambda^{2} \right) \, ,
\end{equation}
which apart from the $\overline{u}$ term is precisely the ( Landau-
gauge ) {\it{perturbative}} one-loop divergence. This is as expected,
since by the very construction of our vertices, (4.32) must contain the
correct perturbative limit. Appealing now once more to our basic
condition (2.5) to demand that this divergence should be removable
by the perturbative counterterm, one finds that the choice
\begin{equation}
\overline{u} = 0
\end{equation}
is dictated uniquely, eliminating the apparent arbitrariness in
the matching condition (4.11). This observation is particularly
interesting in the (non-QCD) context of the even-$r$ "particle" sequence
with a real-axis
propagator pole, since it shows that a nonzero mass term, $u_{r,1}
\Lambda^{2}$, can in principle be generated {\it{while using only the
massless perturbative counterterm}}.
 
The finite term of (4.32) in $R = MS$,
\begin{equation}
\left\{ \left[ \frac{g (\nu)}{4 \pi} \right]^{2} \Gamma^{\left[ \left.
r,1 \right) \right.}_{TT} (k^{2}) \right\}_{MS} =
\left[ \frac{g (\nu)}{4 \pi} \right]^{2} (k^{2})
\left\{ K^{(r)}_{0} (k^{2}) - \left[ \frac{1}{2} N_{C} \left(
\frac{13}{3} \right) - \frac{2}{3} N_{f} \right]
\left(\frac{1}{\epsilon}\right)
\right\}_{\epsilon
\rightarrow 0} \, ,
\end{equation}
where
\begin{equation}
K^{(r)}_{0} (k^{2}) = \frac{J^{(r)}_{0} (k^{2}) - J^{(r)}_{0} (0)}
{k^{2}} \, ,
\end{equation}
represents the renormalized, one-loop, radiative correction, having
errors both of order $g^{4}$ in the perturbative and of type $e (r + 1)$
in the nonperturbative direction.
 
A number of comments are in order here:
\begin{description}
\item[(i)] The divergence content of the one-loop dressing functional is
now exhausted. With $[r,0]$ input, it has just been sufficient for
producing the divergence (4.33) in the perturbative correction, curable
by perturbative means, and for triggering the "eating" mechanism of eq.
(4.28) in the $r+1$ nonperturbative terms, as needed for
self-reproduction of the input. It does not, therefore, generate the
terms of the next higher order $(r + 1)$ automatically, as it does in
the perturbative direction. Even less can it define a unique splitting,
valid uniformly at all $k^{2}$, into order$-g^{0}$ and order$-g^{2}$
terms; this can only be approached asymptotically in the $r >\!> 1$
limit. At finite $r$, output-input matching can be achieved only in a
finite number of comparison data.
\item[(ii)] While perturbative renormalization has now gone through
completely at one loop, one notes that it has done so with a peculiar
new twist: for the vertex function as a whole, the renormalization
is no more multiplicative, as in pure perturbation theory, since the
nonperturbative terms, by (4.28), have established themselves in a
finite manner. At best, one may write the renormalization process as
\begin{equation}
\left(\Gamma_{T} \right)_{R} = \left(Z_{3} \right)_{R} \Gamma^{({\rm
pert})}_{T} + \left(\Gamma_{T} - \Gamma^{({\rm pert})}_{T} \right) \, ,
\end{equation}
with $Z_{3}$ the perturbative gluon-field renormalization constant,
but not as one overall rescaling. (This property must hold already
in the OPE if one wants to avoid nonlocal counterterms.)
\item[(iii)] The observation that the self-consistency mechanism is tied
to the divergences of DS loops immediately implies that nonperturbative
terms of the present type cannot establish themselves in superficially
{\it{convergent}} vertices at one loop. The argument can be extended,
however [7]: an n-loop contribution to a superficially convergent
vertex has prefactor $(g^{2}_{0})^{n}$ but can have at most $n-1$
divergent subintegrations, which through eq. (4.28)
can "eat" at most $n-1$ powers of
$g^{2}_{0}$, so that the result is at best an order-$g^{2}$ correction
but never of order $g^{0}$. This is noteworthy because it shows that,
at least for the limited purpose of determining zeroth-order
nonperturbative parts, the infinitely coupled nature of the DS
system can be beaten without introducing "decoupling" approximations:
{\it{the self-consistency problem of $\Gamma^{[r,0]}$ is strictly
confined to the set of seven DS equations for the superficially
divergent vertices.}}
\item[(iv)] The fact that the one-loop evaluation of $\Phi$ has produced
both terms of order $g^{0}$ and $g^{2}$ demonstrates that there is a
decoupling of the loop and perturbative orders, $l$ and $p$ -- again a
feature
presumably present in any truly nonperturbative solution. A discussion
of this aspect has been given in the first of refs. [7] and need not be
repeated here. Strictly speaking it implies that to fully characterize
an approximant $\Gamma^{[r,p]}$, one would need still another index $l$,
of a technical nature, indicating from what loop order $l \ge p$ the
r.h. sides of the 0-th order self-consistency equations (which become
power series in $\frac{1}{\beta_{0}}$) and the $1 \le p' \le p$
corrections have been determined.
\end{description}
Eqs. (4.30/31) represent, of course, only a small subset of the full
set of equations for the seven vertices (1.14). To establish the full
set of self-consistency conditions by isolating the divergent parts
of all DS interaction terms for these vertices represents, even at the
lowest
nontrivial ($r=1$ and $l=1$) level, a substantial research program, on
which work is in progress. The example of (4.30/31) should however
be sufficient to demonstrate that those divergent parts are considerably
richer in structure than the perturbative ones, and that in this sense
the evaluation with nonperturbatively modified diagram elements,
$\Gamma^{[r,0]}_{sdiv}$, squeezes more dynamical information from
the DS dressing functionals.
\section{Special aspects of the quasiparticle subsequence}
\setcounter{equation}{0}
 
Among the dynamical possibilities opened up by the extended
iterative scheme, the "quasiparticle" subsequence -- with $r$ odd and
only complex-conjugate singularities in the elementary two-point
functions $D_{T}$ and $S_{F}$ -- clearly is the most interesting one for
QCD. Yet its complex singularities seem to raise special questions of
interpretation and of possible violation of physical principles. While
some of these were mentioned in a purely $r = 1$ context in the second
of refs. [7], the crucial ones can be addressed adequately only with the
perspective provided by the full sequential approximation. In the
following we therefore discuss these conceptual questions one by one
under appropriate catchwords.
 
\subsection{Short-lived elementary excitations}
 
The spacetime propagation characteristics implied by the two-point
vertices (2.18) and (A.25) of the quasiparticle subsequence are
exhibited by continuing the corresponding propagators, $D^{[r,0]}_{T}$
and $S^{[r,0]}_{F}$, to Minkowskian $(k_{M})$ space and
Fourier-transforming to Minkowskian $x_{M}$ space (we recall that the
order of these two steps is essential [30] if one is to arrive at the
{\it{time-ordered}} Minkowskian two-point functions). We write results
only for the scalar function $\tilde{D}_{T} (x^{2}_{M})$, since the
qualitative points will be the same for the two invariant functions
of $\tilde{S}_{F} (x_{M})$. Working from (2.23), one has
\begin{equation}
\tilde{D}_{T}^{[r,0]}\left(x^{2}_{M}\right) = \frac{i \Lambda^{2}}{(2
\pi)^{2}} \sum \limits^{(r-1)/2}_{s=0} \, 2 Re \left\{ \left[
\varrho_{r, 4s+2} \sigma_{r, 4s+3} \right] \frac{K_{1} \left( i
\sqrt{\sigma_{r, 4s+3}} \Lambda \sqrt{x^{2}_{M}} \right)}
{i \sqrt{\sigma_{r, 4s+3}} \Lambda \sqrt{x^{2}_{M}}} \right\} \, ,
\end{equation}
where $K_{1}$ is a modified Bessel function. By eq. (2.24), i.e.
by virtue of asymptotic freedom, the small-distance behavior of this
expression is that of a free propagator. At large timelike distances
($x^{0}_{M} \rightarrow \pm \infty$ or $x^{2}_{M} \rightarrow +
\infty$), assuming that $s=0$ refers to the pole pair closest to the
origin of the $k^{2}$ plane, the asymptotic behavior is
\begin{equation}
\tilde{D}^{[r,0]}_{T} (x^{2}_{M}) \rightarrow
\frac{\chi_{r,0} \Lambda^{\frac{1}{2}}}{\left(2 \pi \sqrt{x^{2}_{M}}
\right)^{\frac{3}{2}}} \, e^{-\gamma_{r,0} \sqrt{x^{2}_{M}}}
\cos \left(\omega_{r,0} \sqrt{x^{2}_{M}} + \varphi_{r,0} \right) \, ,
\end{equation}
with real parameters defined by
\begin{equation}
\begin{array}{rl}
\Lambda \sqrt{\sigma_{r,3}} = & \omega_{r,0} - i \gamma_{r,0} = \Lambda
\sqrt{\sigma_{r,1}^{*}} \, ,\\
\varrho_{r,2} i^{-\frac{3}{2}} (\sigma_{r,3})^{\frac{1}{4}} = &
\chi_{r,0} \, e^{-i \varphi_{r,0}} = \left[\varrho_{r,0} i^{\frac{3}{2}}
(\sigma_{r,1})^{\frac{1}{4}} \right]^{*} \, .
\end{array}
\end{equation}
In contrast to the "radiative" behavior of a real-mass propagator at
timelike distances, this function decays exponentially, with an inverse
lifetime, $\tau^{-1} = \gamma_{r,0}$, controlled by the imaginary parts
of the leading $k^{2}$-plane pole pair. Note that the function decreases
in {\it{both}} directions of the time axis. It describes an elementary
excitation of the (transverse) gluon field, $A^{\mu}_{T} (x)|0>$, that
can exist only as part of a microscopically  short-lived intermediate
state.
 
\subsection{Fields without free-field limits}
 
The full Minkowski-space Green functions for $N$ elementary QCD fields,
denoted generically $\varphi_{i}$ with two-point functions
$\tilde{\Delta}_{i}$, of which at least one is a transverse-gluon field
with propagator (5.1), have structure
\begin{equation}
\begin{array}{rl}
\tilde{G}_{N}(x_{1} \ldots x_{N}) \equiv & <0|T \left\{\varphi_{1}
(x_{1})
\ldots A_{T} (x_{k}) \ldots \varphi_{N} (x_{N}) \right\}|0>\\[3ex]
= & \tilde{D}_{T} (x_{k} - x_{l}) \tilde{G}_{N-2} (\ldots) +
\left[\tilde{D}_{T}(x_{k} - x_{l}) \tilde{D}_{T} (x_{m} - x_{n}) +
\right.\\[2ex]
\mbox{} & \left. + \, {\rm{(permutations)}}\right] \, \, \tilde{G}_{N-4}
(\ldots) + \, \,
{\rm{(other \, \, disconnected \, \, terms)}}\\[2ex]
\mbox{} & + \, \int d^{4} y_{1} \ldots \int d^{4} y_{N}
\tilde{\Delta}_{1}
(x_{1} - y_{1}) \ldots \tilde{D}_{T} (x_{k} - y_{k}) \ldots
\tilde{\Delta}_{N} (x_{N} - y_{N})\\[2ex]
\mbox{} & \times \, \tilde{T}_{N} (y_{1} \ldots y_{k} \ldots y_{N}) \, ,
\end{array}
\end{equation}
where $\tilde{T}_{N}$ is the connected and amputated function. For
$x^{0}_{k} \rightarrow \pm \infty$ with the other $x_{i}$ fixed, the
disconnected-gluon terms, if any, vanish exponentially by (5.2). The
behavior of terms with only connected gluons, as exemplified by the
fully connected last term of (5.4), is also dominated by relation (5.2)
for the $\tilde{D}_{T} (x_{k} - y_{k})$ factor, provided the $y_{k}$
integration is sufficiently convergent. The last point involves a
subtlety: as discussed in sect. 3 and the appendix, the nonperturbative
vertices from which $\tilde{T}_{N}$ is built have one real-axis pole
(at $k^{2}_{M} = u_{r,2} \Lambda^{2}$) in the squared momentum of each
external gluon, and such poles would seem to cause insufficiently
convergent, "radiative" behavior of the $\tilde{T}_{N}$ in (5.4) with
respect to $(y_{k})_{M}$. However, as emphasized in connection with
(4.8), the DS self-consistency conditions see to it that propagators
acquire zeroes at the positions of vertex poles, so that the unamputated
function, with propagators on all legs, retains only the {\it{complex}}
momentum-space propagator poles, and therefore overall exponential-decay
factors as in (5.2) for all transverse-gluon legs. The argument
therefore goes through -- the connected term, too, vanishes
exponentially as $x^{0}_{k} \rightarrow \pm \infty$.
 
Now if all Green functions involving an $A_{T} (x)$ field vanish in this
way, then all matrix elements of $A_{T} (x)$ between normalizable states
decay exponentially as $x^{0} \rightarrow \pm \infty$: the field has
{\it{weak limits of zero}} at timelike infinity, in contrast to fields
with stable-particle, real-axis propagator poles that are known to have
nonvanishing weak limits representing free fields [31].
 
A different perspective on the same property results from looking at
the quantity which "normally" -- i.e. for fields with nonvanishing
weak free-field limits -- defines S-matrix elements: the r.h.s. of the
reduction formula. This is obtained from the connected piece of (5.4)
by going back to $k_{M}$ space and conducting a residue search for all
squared momenta approaching real mass-shell values:
\begin{displaymath}
(S - {\rm{1\!l}}) (k_{1}, \ldots, k_{k}, \ldots, k_{N}) \propto
\end{displaymath}
\begin{equation}
\lim_{k^{2}_{1} \rightarrow \mu^{2}_{1}} \ldots \lim_{k^{2}_{k}
\rightarrow \mu^{2}_{k}} \ldots \lim_{k^{2}_{N} \rightarrow \mu^{2}_{N}}
\left\{ (k^{2}_{1} - \mu^{2}_{1}) \ldots (k^{2}_{k} - \mu^{2}_{k})
\ldots (k^{2}_{N} - \mu^{2}_{N}) \times \right.
\end{equation}
\begin{displaymath}
\left. \times \left[ \Delta_{1} (k^{2}_{1}) \ldots D_{T} (k^{2}_{k})
\ldots
\Delta_{N} (k^{2}_{N}) T_{N} (k_{1}, \ldots k_{k}, \ldots k_{N}) \right]
\right\} \, .
\end{displaymath}
In the "normal" case, the full propagators $\Delta_{i}$ have particle
poles at mass shells $k^{2}_{i} = \mu^{2}_{i}$, and the residue search
locks in at those points to give a nonzero S-matrix element,
proportional to the all-on-shell $T$. For a gluon with propagator (5.1),
$D_{T} (k^{2}_{k})$ has only complex pole pairs staying at least a
distance $(Im \sigma_{r,1})\Lambda^{2}$ away from the real $k^{2}_{k}$
axis, and by its numerator zero at $k^{2}_{k} = u_{r,2} \Lambda^{2}$
cancels the real-axis pole of $T$. The residue search at real
$k^{2}_{k}$, no matter what the value of the real $\mu^{2}_{k}$,
therefore gives zero. (The longitudinal gluons do retain their real-axis
poles at $k^{2} = 0$, as do ghosts in the present scheme, but these
unphysical modes are of course excluded from the $S$ matrix by
appropriate projectors not explicitly indicated in (5.5).) One concludes
that S-matrix elements involving at least one external gluon (or, by
extension, quark) are, depending on semantic taste, either nonexistent
(if one considers them as defined only for fields with nonzero
free-field limits) or zero (if one applies definition (5.5) in an
extended sense). The physical implication is the same - gluons and
quarks are not asymptotically detectable reaction products.
 
\subsection{Violation of Causality?}
 
To avoid confusion here, we must recall that arguments deducing absence
of complex singularities in amplitudes from microcausality refer to
{\it{S-matrix elements}} (the propagators themselves, like all off-shell
Green functions, are always "acausal" by construction, i.e. nonvanishing
at spacelike separations) and to those of the {\it{elementary}} fields
(composite hadronic fields, being nonlocal, have no strictly microcausal
commutators). The question then arises whether one-particle reducible
contributions to S-matrix elements, in which one of the external
sums-of-momenta flows through a single gluon or quark propagator,
violate causality through their complex poles. (The simplest amplitudes
of this type are the tree graphs of fig. 3). The answer is no, since
these amplitudes also have external legs corresponding to the elementary
(gluon and quark) fields, and therefore are zero (or must be declared
nonexistent) as discussed in 5.2 above. In other words, the {\it{same}}
set of propagators that seem to raise causality problems when occurring
on internal lines, also ensure the vanishing of those amplitudes {\it
{taken as S-matrix elements}} by
being present on the external lines.
 
\subsection{Reflection Positivity?}
 
In Euclidean-coordinate $(x_{E})$ space, the propagator
$\tilde{D}_{T}^{[r,0]} (x^{2}_{E})$ has an asymptotic form similar to
(5.2) but with the roles of $\omega_{r,0}$ and $\gamma_{r,0}$
interchanged. At $r=1$ and generally for low $r$ in the quasiparticle
sequence, such a propagator is not a purely positive function of
$x^{2}_{E}$, as required by reflection positivity. Its graph, shown
schematically in fig. 4, has negative "overshoots" caused by the
oscillations of $K_{1}$ Bessel functions at complex arguments. This
problem is potentially serious, but not inevitably so. It is quite
possible, and indeed likely (given the fact that the DS equations are
compatible with physical requirements like positivity) that it gets
cured gradually, by the coherent admixture of faster oscillations from
more distant complex-pole pairs, as the order $r$ is increased. (This
in fact is the normal situation in any discrete Fourier approximation of
a positive function.) If this occurs, then the lack of reflection
positivity at low $r$ should simply be viewed as one of the natural
errors expected in the low orders of any approximation scheme. It will
basically be no more serious than e.g. the well-known violation of
unitarity by the Born approximation, i.e. by low-order perturbation
theory. However, it is true that such a conjecture can be checked only
by an actual study of the $r \ge 3$ levels.
 
If present, restoration-by-interference of reflection positivity would
highlight a deeper point about the short-lived elementary excitations:
the single complex-pole pair of the $r = 1$ propagator makes sense,
strictly speaking, only as the lowest level of an improving sequence
of approximations. While a stable-particle pole makes physical sense
in isolation (describing then a free field), a quasiparticle pole pair
can support itself only through the presence of interactions, which
manifest themselves as dressing cuts. It is tempting, though at present
speculative, to view this in parallel with the fact that a nonabelian
gauge field cannot be defined without self-interaction. The formation
of short-lived elementary excitations may be a dynamical specialty of
nonabelian gauge theories.
 
\subsection{Asymptotic Incompleteness}
 
The short-lived gluon or quark excitations as described by a two-point
function in the "quasiqarticle" subsequence are distinct from ordinary
decaying particles or resonances. The difference can be made precise
by looking at the singularity structure of the propagator: for a
resonance, this is characterized by a physical, gauge-fixing independent
cut on the real axis, whose discontinuity shows some local enhancement
around the resonance energy. This indicates that the resonance will
decay into stable, asymptotically detectable fragments, which together
carry the open quantum numbers of that resonance. Therefore, it can in
principle be recreated by macroscopically preparing beams of those
fragments and colliding them. The propagator of the short-lived QCD
excitation has no such decay cut. (Again, there is a subtlety -- these
propagators do have some real-axis cuts, arising from virtual conversion
into longitudinal gluons and ghosts, the unphysical degrees of freedom,
but these are gauge-fixing artefacts recognizable by their $\xi$
dependence.) The excitation therefore cannot decay into asymptotically
detectable configurations carrying its open quantum numbers, and
conversely {\it{cannot be prepared in isolation}} by colliding
macroscopically preparable fragments.
 
What can be produced from asymptotically controlled initial states are
two-gluon and multi-gluon (or multi-quark) configurations compatible
with the quantum numbers of the production channel, whose propagation
in lowest order is described by product propagators of type
\begin{equation}
\left\{ D(k_{1}) D(k_{2}) \ldots D(k_{n}) \right\}^{a} \, ,
\end{equation}
the notation indicating a coupling to the quantum numbers $a$. These
propagators see to it that the configuration propagates only for a
limited time of order $\Lambda^{-1}$ before the dynamics of the theory
forces it to pass its conserved quantities, and share of probability
flux, back to some asymptotically accessible exit channels.
 
The nature of the state space of such a system can at present be
inferred only by educated guess. (Although the state space is in
principle determined by the full set of Green functions, the
prescription to construct the superficially convergent functions by
skeleton expansions is far too inexplicit to actually provide a handle
for an application of the reconstruction theorems.) Fig. 5 is a
schematic representation of such a guess. It depicts the passing of unit
total probability through the continuum space ${\cal{H}}_{A}$ of an
ordinary
scattering system, where it suffers some delay called a resonance, and
through a subspace of exactly conserved quantum numbers, having an
asymptotically accessible portion ${\cal{H}}_{A}$, in QCD.
In the latter case, probability flux can partially and temporarily
-- for times of order $\Lambda^{-1}$ -- leak out of ${\cal{H}}_{A}$
and populate a "closed", not directly accessible space,
${\cal{H}}_{C}$. The presence of the latter implies that ${\cal{H}}_{A}$
does not, together with the bound-state space ${\cal{H}}_{B}$,
exhaust the full state space, i.e. that {\it{such a system lacks
asymptotic completeness}}, a feature unfamiliar but by no means
unacceptable physically. (Since temporal evolution in such a system maps
a smaller onto  a larger space and vice versa, it may have to be
described by operators that are isometric but not unitary, a feature not
excluded by the requirement of probability conservation.)
 
It is natural to inquire about the evolution of a (purely hypothetical,
because of the impossibility of macroscopic preparation) single gluonic
excitation, which would live in a subspace of conserved quantum numbers
of the pure ${\cal{H}}_{C}$ type, without any ${\cal{H}}_{A}$ and
${\cal{H}}_{B}$ portions. This
evolution would exhibit some analogy with that of a free particle in
quantum mechanics, whose wave function, as long as no position
measurement is performed, spreads indefinitely over all space. In a
similar manner, the single short-lived excitation would convert into
more and more complex multiexcitations, all short-lived, and thereby
spread indefinitely over the whole ${\cal{H}}_{C}$ space.
 
\subsection{S-Matrix Unitarity}
 
While evolution projected onto the asymptotically accessible subspace
${\cal{H}}_{A}$ is not unitary at small times, due to the partial
leakout of probability into the ${\cal{H}}_{C}$ space, the fact
that the leaking is dynamically limited to times of order $\Lambda^{-1}$
guarantees that between {\it{asymptotic}} times, $t = -\infty$ and $t' =
+\infty$, probability conservation does hold on that subspace alone,
since all probability flux fed into the system through scattering
eventually gets pushed back into asymptotically accessible exit
channels. In other words, one expects the $S$ matrix to remain strictly
unitary.
 
It is instructive to check this expectation out in a more formal way by
asking whether the relation
\begin{equation}
Im <b|T|a> = - \sum \limits_{n} <b|T^{\dagger}|n> <n|T|a>
\end{equation}
is compatible with the presence of the "closed" subspace
${\cal{H}}_{C}$. If at least one of the two states $|a>, |b>$ is in that
subspace, then by the discussion around eq. (5.5), the l.h.s. and at
least one of the factors on the r.h.s. must be set equal to zero, and
the relation is trivially fulfilled. If both $|a>$ and $|b>$ are in the
asymptotically accessible space ${\cal{H}}_{A}$, then states $|n>$
in ${\cal{H}}_{C}$, again by the arguments of sect. 5.2, make no
contribution to
the sum on the r.h.s. But then the imaginary part on the l.h.s., which
is equal to the discontinuity across the timelike real axis of the
relevant energy-squared variable for the $a \rightarrow b$ process,
does not receive contributions from such states either: intermediate
states in ${\cal{H}}_{C}$ are described by propagators of type (5.6),
and integration over them produces {\it{pairs}} of cuts at
complex-conjugate locations, which together always leave the amplitude
continuous and real along the real $s$ axis (even though some of them
do start at points on that axis), i. e. which do not produce physical
absorptive parts. ( This can be checked out in detail for e. g. the
one-loop terms of fig. 2 with the $r = 1$ propagators ). One concludes
that it is indeed consistent for eq. (5.7) to hold on the asymptotically
accessible space alone.
 
The discussion of this section may be summarized by stating that,
perhaps contrary to first appearances, a solution $\{\Gamma^{[r,p]}|r\,
\, {\rm{odd}}\}$ of the quasiparticle type with complex-conjugate
propagator singularities does not necessarily imply violation of
physical principles beyond the natural errors expected in the lower
orders of any systematic approximation scheme. Such a solution is in
fact well suited for a description of the elementary QCD excitations
where they come closest to being "seen", namely at the origins of jet
events. \\
\vspace{4cm}
 
\begin{center}
{\large \bf{Acknowledgment}}
\end{center}
Parts of this paper were written during a visit of the author to the
Department of Particle Physics of the Weizmann Institute of Science,
Rehovot, Israel. The author is grateful to Professors Y. Frishman and
M.W. Kirson for the hospitality extended to him at the Department and
for the support he received through the Einstein Center of
Theoretical Physics.
\newpage
\begin{center}
{\large \bf{References}}
\end{center}
\begin{description}
\item[ [1] ] J.C. Le Guillou and J. Zinn-Justin (eds.), Large-Order
Behaviour of Perturbation Theory, North-Holland, Amsterdam, 1990
\item[ [2] ] G. t'Hooft, in: A. Zichichi (Ed.), The Whys of Subnuclear
Physics (Proceedings Erice 1977), Plenum, New York, 1979
\item[ [3] ] D.J. Gross and A. Neveu, Phys. Rev. {\bf{D10}} (1974) 3235;
C.G. Callan, R.F. Dashen, and D.J. Gross, Phys. Rev. {\bf{D17}} (1978)
2717
\item[ [4] ] N.K. Nielsen, Nucl. Phys. {\bf{B120}} (1977) 212; J.C.
Collins, A. Duncan, and S.D. Joglekar, Phys. Rev. {\bf{D16}} (1977) 438
\item[ [5] ] G. M\"unster, Z. Physik {\bf{C12}} (1982) 43
\item[ [6] ] A short account of some of the ideas in this paper has
appeared in: K. Goeke, H.Y. Pauchy Hwang and J. Speth (Editors), Medium
Energy Physics (Proceedings German-Chinese Symposium, Bochum 1992),
Plenum, New York, 1994.
\item[ [7] ] U. H\"abel, R. K\"onning, H.-G. Reusch, M. Stingl and S.
Wigard, Z. Physik {\bf{A336}} (1990) 423 and 435
\item[ [8] ] F.J. Dyson, Phys. Rev. {\bf{75}} (1949) 1736; J. Schwinger,
Proc. Nat. Acad. Sci. {\bf{37}} (1951) 452 and 455
\item[ [9] ] E.J. Eichten and F.L. Feinberg, Phys. Rev. {\bf{D10}}
(1974) 3254
\item[ [10] ] E.g. T. Muta, Foundations of Quantum Chromodynamics, World
Scientific, Singapore, 1987
\item[ [11] ] E. g. P. Pascual and R. Tarrach, QCD: Renormalization for
the Practitioner (Lecture Notes in Physics Vol. 194), Springer, Berlin,
1984
\item[ [12] ] D. Zwanziger, Nucl. Phys. {\bf{B323}} (1989) 513
\item[ [13] ] D. Zwanziger, Nucl. Phys. {\bf{B378}} (1992) 525, ibid.
{\bf{B399}} (1993) 477
\item[ [14] ] V.N. Gribov, Nucl. Phys. {\bf{B139}} (1978) 1
\item[ [15] ] N. Maggiore and M. Schaden, Phys. Rev. {\bf{D50}} (1994)
6616
\item[ [16] ] E.g. C.-R. Ji, A.F. Sill, and R.M. Lombard-Nelson, Phys.
Rev. {\bf{D36}} (1987) 165
\item[ [17] ] M. L\"uscher, R. Sommer, P. Weisz and U. Wolff, Nucl. Phys.
{\bf{B389}} (1993) 247, ibid. {\bf{B413}} (1994) 481
\item[ [18] ] V.N. Gribov, Physica Scripta {\bf{T15}} (1987) 164
\item[ [19] ] H. Lehmann, Nuovo Cim. {\bf{11}} (1954) 342,
G. K\"allen, Helv. Phys. Acta {\bf{25}} (1952) 417
\item[ [20] ] E.g. G.A. Baker, Jr., Essentials of Pad$\acute{\rm{e}}$
Approximants, Academic Press, New York/London, 1975
\item[ [21] ] J. Schwinger, Phys. Rev. {\bf{125}} (1962) 397
\item[ [22] ] M. Stingl, Phys. Rev. {\bf{D34}} (1986) 3863; Erratum,
ibid. {\bf{D36}} (1987) 651
\item[ [23] ] G. Barton, Introduction to Advanced Field Theory,
Interscience, New York, 1963
\item[ [24] ] M. Lavelle and M. Schaden, Phys. Lett. {\bf{B208}} (1988)
297
\item[ [25] ] M. Lavelle and M. Oleszczuk, Z. Physik {\bf{C51}} (1991)
615
\item[ [26] ] J. Ahlbach, M. Lavelle, M. Schaden and A. Streibl, Phys.
Lett. {\bf{B275}} (1992) 124
\item[ [27] ] M. Lavelle and M. Oleszczuk, Mod. Phys. Lett. {\bf{A7}}
(1992) 3617 (Brief Reviews)
\item[ [28] ] J.S. Ball and T.-W. Chiu, Phys. Rev. {\bf{D22}} (1980)
2550
\item[ [29] ] G. t'Hooft, Nucl. Phys. {\bf{B61}} (1973) 455
\item[ [30] ] E. g. G. Roepstorff, Path-Integral Approach to Quantum
Physics, Springer, New York, 1994, ch. 7.2
\item[ [31] ] H. Lehmann, K. Symanzik, and W. Zimmermann, Nuovo Cim.
{\bf{1}} (1955) 1425
\item[ [32] ] J.S. Ball and T.-W. Chiu, Phys. Rev. {\bf{D22}} (1980)
2542
\end{description}
\newpage
\begin{center}
\Large\bf{Appendix}
\end{center}
\renewcommand{\theequation}{\Alph{section}.\arabic{equation}}
\setcounter{equation}{0}
\begin{appendix}
\section{Other Superficially Divergent Vertices}
\subsection{Four-Vector Vertex}
This is the highest of the superficially divergent vertices (1.14).
Its complexity with respect
to color and Lorentz structure exceeds by an order of magnitude that
of all the other $\Gamma_{sdiv}$ combined, and is the main source of
technical difficulty in the $p = 0$ self-consistency problem. The number
of color-basis tensors $C^{(i)}$ -- each a fourth-rank object over the
adjoint representation of $SU(N_{C})$ -- in the decomposition
\begin{equation}
\Gamma^{\kappa \lambda \mu \nu}_{4V, abcd} = \sum \limits_{i} \,
C^{(i)}_{abcd} \, \, \, \Gamma^{\kappa \lambda \mu \nu}_{4V (i)}
\end{equation}
is eight for $N_{C} = 3$ and nine for $N_{C} \ge 4$. For brevity, we
refer the reader to ref. [11] for the general case and write an
example of a suitable color basis only for the case of immediate
interest in QCD, viz. $N_{C} = 3$:
\begin{equation}
\begin{array}{r@{=}l}
C^{(1)}_{abcd} & \delta_{ab} \delta_{cd} \, , \, \, C^{(2)}_{abcd} =
\delta_{ac} \delta_{bd} \, ,
\, \, C^{(3)}_{abcd} = \delta_{ad} \delta_{bc} \, ,\\[4ex]
C^{(4)}_{abcd} & f_{abn} f_{cdn} \, , \, \, \left[C^{(5)} - C^{(6)}
\right]_{abcd} = f_{acn} f_{dbn} - f_{adn} f_{bcn} \, ,\\[4ex]
C^{(7)}_{abcd} & d_{abn} f_{cdn} \, , \, \, C^{(8)}_{abcd} =
d_{acn} f_{dbn} \, ,
\, \, C^{(9)}_{abcd} = d_{adn} f_{bcn} \, .\\
\end{array}
\end{equation}
The combination $C^{(5)} + C^{(6)}$ is not listed, as it equals
$- C^{(4)}$ by virtue of the Jacobi identity. In the many situations
where minimality of the basis is not crucial but manifest Bose
and crossing symmetry are, we shall nevertheless use the
nine-dimensional set including $C^{(5)}$ and $C^{(6)}$ separately.
 
For Lorentz structure, we again concentrate on the totally transverse
pieces defined in analogy with (3.2), which have tensor decompositions
\begin{equation}
\begin{array}{rl}
\Gamma^{\kappa' \lambda' \mu' \nu'}_{4T (i)} (p_{1} \ldots p_{4}) = &
t^{\kappa' \kappa} (p_{1}) t^{\lambda' \lambda} (p_{2}) t^{\mu' \mu}
(p_{3}) t^{\nu' \nu} (p_{4})\\[3ex]
\mbox{} & \times \sum \limits^{2}_{k=0} \, \sum
\limits_{j} L^{\kappa \lambda \mu \nu}_{(k,j)} (p_{1} \ldots p_{4})
\, \, \, G_{(i) k, j} \left(p^{2}_{1} \ldots p^{2}_{4}; s, t, u \right)
\,
,
\end{array}
\end{equation}
with $\sum p_{{i}} = 0$ again understood. Linearly independent Lorentz
tensors $L_{(k,j)}$, constructed from the Euclidean metric
$\delta^{\alpha \beta}$ and three independent combinations (denoted
generically $k_{1, 2, 3}$) of the $p_{i}$, come with three mass
dimensions, $2k = 0,2$, and $4$. They include three dimensionless $(k =
0)$ tensors,
\begin{equation}
L^{\kappa \lambda \mu \nu}_{(0,1)} = \delta^{\kappa \lambda} \delta^{\mu
\nu} \, , \, \,
L^{\kappa \lambda \mu \nu}_{(0,2)} = \delta^{\kappa \mu} \delta^{\lambda
\nu} \, , \, \,
L^{\kappa \lambda \mu \nu}_{(0,3)} = \delta^{\kappa \nu} \delta^{\lambda
\mu} \, ,
\end{equation}
plus 54 independent $2k = 2$ tensors $L_{(1,j)}$ of type $\delta^{\alpha
\beta} k^{\gamma}_{m} k^{\delta}_{n} \quad(m, n \in \{1, 2, 3\})$, of
which
24 contribute to the totally transverse vertex, plus 81 independent
$2k = 4$ tensors $L_{(2,j)}$ of type $k^{\kappa}_{k} k^{\lambda}_{l}
k^{\mu}_{m} k^{\nu}_{n} \quad(k, l, m, n \in \{1, 2, 3\})$, of which
only
16 contribute to the $4T$ vertex. The perturbative zeroth-order vertex
contains only the $k = 0$ Lorentz tensors (A.4) and has the expressions
\begin{equation}
\begin{array}{r@{=}l}
\Gamma^{(0){\rm{pert}}}_{4V} & C^{(4)} \left[ L_{(0,2)} - L_{(0,3)}
\right] + C^{(5)} \left[ L_{(0,3)} - L_{(0,1)} \right] + C^{(6)}
\left[ L_{(0,1)} - L_{(0,2)} \right]\\[4ex]
\mbox{} & C^{(4)} \left[ \frac{3}{2} \left(L_{(0,2)} - L_{(0,3)} \right)
\right] + \frac{1}{2} \left( C^{(5)} - C^{(6)} \right) \left[
-2L_{(0,1)} + L_{(0,2)} + L_{(0,3)} \right]
\end{array}
\end{equation}
in the linearly dependent but crossing symmetric and in the minimal
color basis (A.2), respectively. We do not write full enumerations of
the $k > 0$ objects here, which will be of interest only in connection
with detailed loop computations. For a first exploration of the solution
described here, it will be reasonable in any case to start by looking
for approximate self-consistency with a strongly restricted tensor
structure.
 
One would expect to treat the invariant functions $G_{(i) k, j}$
associated with these tensors as functions of six independent
Lorentz-invariant variables, an example being the six scalar products
formed from the three conserved total four-momenta in the three crossed
two-gluon channels,
\begin{equation}
\begin{array}{r@{=}l@{=}l}
P \, \, & \, \, p_{1} + p_{2} \, \, & \, \, -(p_{3} + p_{4}) \, ,\\
R \, \, & \, \, p_{1} + p_{3} \, \, & \, \, -(p_{2} + p_{4}) \, ,\\
Q \, \, & \, \, p_{1} + p_{4} \, \, & \, \, -(p_{2} + p_{3}) \, .
\end{array}
\end{equation}
Again, however, it turns out that one has no freedom here: dynamical
consistency in the three-gluon DS equation connecting $\Gamma_{3V}$
to $\Gamma_{4V}$, together with crossing symmetry, dictates that
rational approximation of $G's$ be performed with respect to the
{\it{seven}} variables indicated in (A.3), where
\begin{equation}
s = P^{2} \, , \qquad u = R^{2} \, , \qquad t = Q^{2} \, ,
\end{equation}
and with the relation expressing their linear dependence,
\begin{equation}
s + u + t = \sum \limits^{4}_{i=1} p^{2}_{i} \, ,
\end{equation}
being carried as a subsidiary condition. This would seem to
considerably
complicate the writing and proper restriction of FDRA's were it not for
two simplifying features which we anticipate: the various additive terms
of (A.1/A.3) can be grouped as either crossing triplets or crossing
singlets, and if one restricts attention to $\Gamma^{[r,0]}_{4T}$, the
nonperturbatively modified vertex of zeroth perturbative order, then
self-consistency is possible for a simplified structure, where in each
member of a crossing triplet the invariant-function {\it{denominator}}
contains only one of the Mandelstam variables $s, t, u$ at a time, while
for a crossing singlet it does not depend on them at all. Moreover, pole
positions in FDRA's with respect to the four $p^{2}_{i}$ and with
respect to the variables (A.7), which a priori could be chosen
differently, are forced by DS consistency to be in fact the same. Thus
the invariant function for the s-channel member of a crossing triplet
would possess r-th degree FDRA's of the form
\begin{equation}
G^{[r,0]}_{(i)k, j} = \frac{N^{(r)}_{4V (i)k, j} (p^{2}_{1} \ldots
p^{2}_{4}; s; \frac{1}{2} (u - t))}
{\left[ \prod \limits^{r}_{\sigma = 1} \left( s + u^{''}_{r, 2 \sigma}
\Lambda^{2} \right) \right] \left\{ \prod \limits^{4}_{l=1} \left[ \prod
\limits^{r}_{\sigma = 1} \left( p^{2}_{l} + u^{''}_{r, 2 \sigma}
\Lambda^{2} \right) \right] \right\}} \, ,
\end{equation}
and the other two members of the triplet would follow from this by the
two crossing operations
\begin{equation}
\begin{array}{rl}
(b, \lambda, p_{2}) \longleftrightarrow (c, \mu, p_{3}) ; &
P \leftrightarrow R , s \leftrightarrow u \, ;\\
(b, \lambda, p_{2}) \longleftrightarrow (d, \nu, p_{4}) ; &
P \leftrightarrow Q , s \leftrightarrow t \, .
\end{array}
\end{equation}
The mass dimension $-2k$, as well as the various restrictions and
Bose-symmetry properties, are built into the numerator polynomial,
\begin{equation}
\begin{array}{ll}
N^{(r)}_{4V(i)k,j} \left(p^{2}_{1} \ldots p^{2}_{4}; s; \frac{1}{2}
(u-t) \right) = & \left( \delta_{i4} + \delta_{i5} + \delta_{i6} \right)
\delta_{k0} \left( s p^{2}_{1} p^{2}_{2} p^{2}_{3} p^{2}_{4}
\right)^{r}\\[3ex]
\mbox{} & {\hspace{-4.5cm}}+ \sum \limits_{m_{1}\ldots n_{2} \ge 0}
C^{(i)k,j(r)}_{m_{1}
m_{2} m_{3} m_{4} n_{1} n_{2}} \left[ \prod \limits^{4}_{l=1}
(p^{2}_{l})^{m_{l}} \, \, s^{n_{1}} \left( \frac{u-t}{2} \right)^{n_{2}}
\right] (\Lambda^{2})^{5r-k-(m_{1}+ \ldots + n_{2})} \, .
\end{array}
\end{equation}
Condition (2.3) allows nonzero coefficients $C_{m_{1} \ldots n_{2}}$
only for
\begin{equation}
m_{1} + m_{2} + m_{3} + m_{4} + n_{1} + n_{2} \le 5r - k - 1 \, .
\end{equation}
The additional restrictions arising from condition (2.5) are now
more involved: when $\Gamma_{4V}$ appears in a 1PI diagram, up to
{\it{two}} of its four legs may be external. To avoid the occurrence
of ultraviolet divergences stronger than the perturbative ones,
no term of the vertex should then behave worse than a constant
at large loop momenta when (i) any three of its momenta $p_{i}$ are
running in loops and (ii) any two of its momenta are running in a loop.
Considering only the case where the $2k$ momenta supplied (for $k > 0$)
by the $L_{(k,j)}$ tensor are all running along in those loops, we find
that (i) imposes the four restrictions
\begin{equation}
\left[ \sum \limits^{4}_{l=1} (1 - \delta_{li}) m_{l} \right] +
(n_{1} + n_{2}) \le 4r - k \qquad (i = 1, 2, 3, 4) \, ,
\end{equation}
whereas (ii) implies, for the function (A.11), the six restrictions
\begin{equation}
\begin{array}{rl}
(m_{i} + m_{j}) + (n_{1} + n_{2}) & \le \, \, 3r - k\\
(i,j) & = (1,3), (1,4), (2,3), (2,4) \, ;
\end{array}
\end{equation}
\begin{equation}
\begin{array}{r@{\le}l}
2 (m_{1} + m_{2}) + n_{2} \, \, & \, \, 4r - 2k \, ;\\[4ex]
2 (m_{3} + m_{4}) + n_{2} \, \, & \, \, 4r - 2k \, .
\end{array}
\end{equation}
In writing (A.15) we have used the fact that
\begin{equation}
u - t = (p_{2} - p_{1}) \cdot (p_{4} - p_{3}) \, .
\end{equation}
In other cases where some of the "tensorial" momenta stay out of
the loops, these restrictions may be relaxed slightly, but the
additional freedom gained in this way is minor.
 
For the self-consistency problem, the relevant structural property
emerging
from the above is that with respect to the variables (A.7), the $p = 0$
vertex may be assumed to have a simplified partial-fraction
decomposition which is the analog of (3.10),
\begin{equation}
\begin{array}{rl}
\Gamma^{[r,0]}_{4T} = & E^{(r)}_{0} + \sum \limits^{r}_{n=1} \left[
E^{(r)}_{n(s)} \left( \frac{\Lambda^{2}}{s + u^{''}_{r,2n} \Lambda^{2}}
\right) + \right.\\[4ex]
\mbox{} & + \left. E^{(r)}_{n(u)} \left ( \frac{\Lambda^{2}}{u +
u^{''}_{r,2n}
\Lambda^{2}} \right) + E^{(r)}_{n(t)} \left( \frac{\Lambda^{2}}{t +
u^{''}_{r,2n} \Lambda^{2}} \right) \right] \, .
\end{array}
\end{equation}
Here $E_{n(u)}, E_{n(t)}$ are the crossing partners of $E_{n (s)}$ in
the sense of (A.10), and each of the $E_{n}$ tensors in turn has
a partial-fraction decomposition with respect to the $p^{2}_{i}$
variables whose form is exemplified by
\begin{equation}
E^{(r)}_{n(s)} = \sum \limits^{1}_{\alpha, \beta = 0} \sum
\limits^{\alpha r}_{\sigma = \alpha} \sum \limits^{\beta r}_{\tau =
\beta} \, E_{n(s) \sigma \tau} \left( \frac{\Lambda^{2}}{p^{2}_{1} +
u''_{r,2 \sigma} \Lambda^{2}} \right)^{\alpha} \left(
\frac{\Lambda^{2}}{p^{2}_{2} + u''_{r, 2 \tau} \Lambda^{2}}
\right)^{\beta} \, .
\end{equation}
Here the $E_{n (s) \sigma \tau}$ functions still have
dependence on $p^{2}_{3}, p^{2}_{4}$.
 
\subsection{Ghost vertices}
The well-known fact that diagrams,
whether perturbative or dressed, with external $G$ and $\overline{G}$
lines factor out one power of an external momentum per $G{\overline{G}}$
pair, causes the Dyson-Schwinger equation (1.13) for the inverse ghost
propagator to assume the special form
\begin{equation}
\delta_{ab} \Gamma_{G{\overline{G}}} (p^{2}) = -\delta_{ab} p^{2}
\left\{ 1 + \left( \frac{g_{0}}{4 \pi} \right)^{2} \tilde{I} \left[
\tilde{D}, D, \Gamma_{GV{\overline{G}}} \right] (p^{2}) \right\} \, ,
\end{equation}
where the dimensionless, logarithmically divergent loop integral
$\tilde{I} (p^{2})$ involves the $\Gamma_{GV{\overline{G}}}$ three-point
vertex. Rational approximants of the odd-r sequence for the ghost
propagator $\tilde{D}$ are therefore of a type we omitted as exotic
when discussing $D_{T}$ -- they generically have two different real-axis
poles, one massless and one massive, which for an unphysical excitation
is not impossible. For example, the counterpart of (2.15) reads,
\begin{equation}
\tilde{D}^{[1,0]} (p^{2}, \Lambda^{2}) =  \frac{p^{2} + v_{1,2}
\Lambda^{2}}{p^{2} (p^{2} + v_{1,1} \Lambda^{2})} \, .
\end{equation}
This includes the special cases $v_{1,2} = v_{1,1}$, where the
$r=1$ function reduces to an  $r=0$ function with one massless
real-axis
pole, $v_{1,2} = 0$, where the same happens but with a mass (generally
gauge-fixing dependent) for the FP ghost, and $v_{1,1} = 0$, where the
ghost develops a second-order pole at $p^{2} = 0$, as it does under the
Gribov-Zwanziger mechanism [14,12]. For a generic gauge fixing,
however,
(A.20) and its higher-r extensions are likely to remain the relevant
forms.
 
In the $\Gamma_{GV{\overline{G}}}$ three-point vertex, FDRA's for the
invariant functions $\tilde{F}_{(c)i} (c = f$ or $d)$ defined by the
tensor decomposition
\begin{equation}
\Gamma^{\hspace{0.2cm}\mu}_{abc} (p_{1}, k, -p_{2})_{GV{\overline{G}}}
= f_{abc}
\left[ p^{\mu}_{1} \tilde{F}_{(f)0} + k^{\mu} \tilde{F}_{(f) 1} \right]
+ d_{abc} \left[ p^{\mu}_{1} \tilde{F}_{(d)0} + k^{\mu} \tilde{F}_{(d)1}
\right] \, ,
\end{equation}
(where now $p_{1} + k - p_{2} = 0$), are somewhat more complicated than
(3.6) because of the absence of symmetry properties. In particular,
denominator zeroes in the ghost-line variables $p^{2}_{1}$ and
$p^{2}_{2}$, which we denote as -- $v'_{r, 2s} \Lambda^{2}$, may differ
from those in the gluon variable $k^{2}$ even at $r = 1$. The general
structure is now
\begin{equation}
\begin{array}{rc}
\tilde{F}^{[r,0]}_{(c)i} = & \frac{N^{(r)}_{GV{\overline{G}} (c) i}
(p^{2}_{1}, k^{2}, p^{2}_{2})}
{ \left[ \prod \limits^{r}_{s=1} \left(p^{2}_{1} + v'_{r,2s} \Lambda^{2}
\right) \right] \left[ \prod \limits^{r}_{s=1} \left( k^{2} +
\tilde{u}'_{r,2s} \Lambda^{2} \right) \right] \left[ \prod
\limits^{r}_{s=1} \left(p^{2}_{2} + v'_{r,2s} \Lambda^{2} \right)
\right]}\\
\mbox{} &
(c = f \, \, {\rm{or}} \, \, d, \quad
i = 0 \, \, {\rm{or}} \, \, 1, \quad
r = 1, 3, 5 \ldots) \, ,
\end{array}
\end{equation}
where numerator polynomials must respect the perturbative limit
\begin{displaymath}
\tilde{F}^{(0){\rm{pert}}}_{(c)i} = \delta_{cf} \delta_{i0} \, ,
\end{displaymath}
as well as restrictions analogous to the $k = 0$ version of (3.8/3.9),
but no symmetry restrictions.
The explicit form of the $\tilde{F}_{(f)0}$ function for $r = 1$ will
again be given for illustration:
\begin{equation}
\begin{array}{rl}
\tilde{F}^{[1,0]}_{(f)0} \left(p^{2}_{1}, k^{2}, p^{2}_{2} \right) = &
1 + y_{1} \tilde{\Pi}_{p_{1}} + y_{3} \tilde{\Pi}_{p_{2}} + \\
\mbox{} & + \left( y_{6} + y_{7} \frac{k^{2}}{\Lambda^{2}} \right)
\tilde{\Pi}_{p_{1}} \tilde{\Pi}_{p_{2}} + \Pi_{k} \left[ y_{2} +
\left( y_{4} + y_{5} \frac{p^{2}_{1}}{\Lambda^{2}} \right)
\tilde{\Pi}_{p_{2}}\right.\\
\mbox{} & \left. + \left( y_{8} + y_{9} \frac{p^{2}_{2}}{\Lambda^{2}}
\right) \tilde{\Pi}_{p_{1}} + y_{10} \tilde{\Pi}_{p_{1}}
\tilde{\Pi}_{p_{2}} \right]
\end{array}
\end{equation}
Here $\tilde{\Pi}_{p} = \Lambda^{2}/\left(p^{2} + v'_{2}
\Lambda^{2}\right)$ denotes a rational building block analogous to the
$\Pi$ of eq. (3.11).
The important
point for the DS self-consistency problem is again the existence of a
"regular-plus-singular" decomposition, analogous to (3.10), in any one
variable; e.g.,
\begin{equation}
\begin{array}{rl}
\left[ \Gamma^{\hspace{0.25cm}\mu}_{GV{\overline{G}}(c)} (p_{1}, k,
-p_{2})
\right]^{[r,0]} = & \tilde{B}^{\mu}_{(c)0,r} \left(p^{2}_{1},
p^{2}_{2};k^{2}\right)\\
\mbox{} & + \sum \limits^{r}_{s=1} \, \tilde{B}^{\mu}_{(c)s,r}
\left(p^{2}_{1}, p^{2}_{2}\right) \left( \frac{\Lambda^{2}}{k^{2} +
\tilde{u}'_{r,2s} \Lambda^{2}} \right) \, ,
\end{array}
\end{equation}
 
In refs [7], using heuristically what we would now call the $r = 1$
degree of rational approximation, treatment of the two ghost vertices
was subject to the prejudice, taken over from older work of Eichten
and Feinberg [9], that vertices do not develop nonperturbative
corrections with respect to the momenta of unphysical external lines.
This, in particular, kept the ghost propagator $\tilde{D}^{[1,0]}$ in
its perturbative form, $1/p^{2}$. From the more systematic viewpoint
advanced here, this prejudice is seen to be unjustified. Even if
DS dynamics would make the case $v_{r,2} = v_{r,1}$ prevail for all
$r$,
the function $\tilde{D}^{[r,0]}$, which in contrast to the longitudinal
gluon propagator $D_{L}$ is not protected by an ST identity, would
pick up pole-zero pairs at $ r \ge 3$ to represent its dressing cuts,
and these in turn would demand $p^{2}_{1}$ and $p^{2}_{2}$ denominator
factors in (A.22). This is in accord with the fact, emphasized by
Lavelle and Schaden [24] in an OPE context, that nonperturbative
ghost-antighost vacuum condensates of zeroth perturbative order do
exist. \\
 
\subsection{Fermion propagators.}
We use Euclidean
$\gamma$
matrices obeying $\{ \gamma^{\mu}, \gamma^{\nu}\} = - 2 \delta^{\mu
\nu}$, so that ${p \hspace{-0.2cm}/}{p \hspace{-0.2cm}/} = -p^{2}$.
The flavor-F, inverse propagator of zeroth perturbative order, analogous
to (2.18), has odd-r approximants
\begin{equation}
- \Gamma^{[r,0]}_{F \overline{F}} ({p \hspace{-0.2cm} /}, \hat{m}_{F},
\Lambda) = {p \hspace{-0.2cm} /} + \kappa^{(F)}_{r,1} + \frac{ \left(
\kappa_{r,3}^{(F)} \right)^{2}}{{p \hspace{-0.2cm} /} +
\kappa^{(F)}_{r,2}}
+ \sum \limits^{(r+1)/2}_{s=2} \left[ \frac{ \left(
\kappa^{(F)}_{r,2s+1} \right)^{2}}{{p \hspace{-0.2cm} /} +
\kappa^{(F)}_{r,2s}} + \frac{\left( \kappa^{* (F)}_{r, 2s+1}
\right)^{2}}{{p \hspace{-0.2cm} /} + \kappa^{(F)*}_{r, 2s}} \right] \, .
\end{equation}
Here we encounter a complicating feature:
the nonperturbative mass scales $\kappa_{r,i}$ cannot, except
in special cases, be written simply as numerical multiples of $\Lambda$,
since there now exist additional RG-invariant mass scales. These are the
extraneous, or Lagrangian, mass scales $\hat{m}_{F}$, one for each
flavor $F$, that are connected to renormalized quark masses $m_{F}
(\nu)$ through
\begin{equation}
\left(\hat{m}_{F} \right)_{R} = \left[ m_{F} (\nu) \right]_{R}
\exp \left\{ - \int \limits^{g(\nu)} \, dg' \left[ \frac
{\gamma_{m} (g')}{\beta(g')} \right]_{R} \right\}
\end{equation}
in the scheme indicated by $R$. The rational character of approximants
with respect to $\Lambda$ then allows fermionic mass scales, such
as the $\kappa$'s and $\kappa^{2}$'s of (A.25), to be polynomials in
$\Lambda$ and $\hat{m}_{f}$. The only restrictions come from condition
(2.3), which demands that the bare vertex,
\begin{equation}
-\Gamma^{(0){\rm{pert}}}_{F{\overline{F}}} = p{\hspace{-0.2cm}/} +
\hat{m}_{F} \, ,
\end{equation}
be approached in the "perturbative" limit, $\Lambda  \rightarrow 0$.
In general, we then have
\begin{equation}
\kappa^{(F)}_{r,1} = \hat{m}_{f} + w^{(F)}_{r,1} \Lambda \, ,
\end{equation}
\begin{equation}
\kappa^{(F)}_{r,2s} = w^{'(F)}_{r,2s} \hat{m}_{f} + w^{(F)}_{r,2s}
\Lambda \, ,
\end{equation}
\begin{equation}
\left(\kappa^{(F)}_{r, 2s+1} \right)^{2} = \Lambda \left[ w^{'(F)}_{r,
2s+1} \hat{m}_{f} + w^{(F)}_{r, 2s+1} \Lambda \right] \, .
\end{equation}
The primed coefficients, $w'$, are absent in the strict chiral limit,
$\hat{m}_{F} = 0$. In the opposite limit, $\hat{m}_{F} >\!> \Lambda$,
the pole terms of (A. 25) are of order $\Lambda/\hat{m}_{f}$ relative
to the leading $\hat{m}_{f}$ term of (A. 27).
To illustrate (A.25), we write the analogs of (2.13) and (2.15) for
$r=1$ (omitting now $F$ on mass scales):
\begin{equation}
\begin{array}{rl}
- \Gamma^{[1,0]}_{F \overline{F}}  & = {p \hspace{-0.2cm}/} +
\kappa_{1,1} + \frac{\kappa^{2}_{1,3}}{{p \hspace{-0.2cm}/} +
\kappa_{1,2}}\\[3ex]
\mbox{} & = {p \hspace{-0.2cm}/} \left[ 1 -
\frac{\kappa^{2}_{1,3}}{p^{2} +
\kappa^{2}_{1,2}} \right] + {\rm{1\!I}} \left[ \kappa_{1,1} +
\frac{\kappa_{1,2} \kappa^{2}_{1,3}}{p^{2} + \kappa^{2}_{1,2}} \right]
\, ,
\end{array}
\end{equation}
\begin{equation}
\begin{array}{rl}
S^{[1,0]}_{F} & = \frac{{p \hspace{-0.15cm}/} + \kappa_{1,2}}{({p
\hspace{-0.15cm}/} + \kappa_{1,+}) ({p \hspace{-0.15cm}/} +
\kappa_{1,-})}\\[3ex]
\mbox{} & = \frac{{-p \hspace{-0.15cm}/} \left[p^{2} + (\kappa_{1,2}
(\kappa_{1,+} + \kappa_{1,-}) - |\kappa_{1, \pm}|^{2}) \right] +
{\rm{1\!I}}
\left[ (\kappa_{1,+} + \kappa_{1,-} - \kappa_{1,2}) p^{2} + \kappa_{1,2}
| \kappa_{1, \pm}|^{2} \right]}{(p^{2} + \kappa^{2}_{1,+}) (p^{2} +
\kappa^{2}_{1,-})}\, ,
\end{array}
\end{equation}
where ${\rm{1\!I}}$ denotes a unit Dirac  matrix, and where
\begin{equation}
\kappa^{2}_{1,\pm} = \left[ \frac{1}{2} \left( \kappa^{2}_{1,1} +
\kappa^{2}_{1,2} \right) - \kappa^{2}_{1,3} \right] \pm i \left(
\kappa_{1,1} + \kappa_{1,2} \right) \sqrt{\kappa^{2}_{1,3} -
\left[\frac{1}{2} \left( \kappa_{1,1} - \kappa_{1,2} \right)
\right]^{2}}
\, .
\end{equation}
A complex-conjugate pair of poles, and therefore a short-lived
quark-like excitation, is present if
\begin{equation}
\kappa_{1,1} + \kappa_{1,2} \not= 0 \, \, \, {\rm{and}} \, \, \,
\kappa^{2}_{1,3} > \left[ \frac{1}{2} \left( \kappa_{1,1} - \kappa_{1,2}
\right) \right]^{2} \, .
\end{equation}
Again, for the interpretation to hold at $r \ge 3$ requires a definite
pattern of one nearest-to-origin pole pair, plus well separated
zero-pole strings identifiable as complex cuts, to emerge as $r$ is
increased.
 
It is interesting to note the tight correlation present in (A.31)
between spontaneous chiral-symmetry breaking and the emergence of
short-lived quarklike excitations. In the chiral limit $\hat{m}_{f} =
0$, where the $\kappa_{1,i}$ scales are purely spontaneous (multiples of
$\Lambda$), chiral-symmetry breaking can be avoided only by having
the invariant function associated with the Dirac $1\!{\rm{l}}$ in (A.31)
vanish identically, which implies either $\kappa_{1,1} = \kappa_{1,2} =
0$, or $\kappa_{1,1} = \kappa^{2}_{1,3} = 0$. In both cases, the pole
positions (A.33) turn purely real. In other words, whenever short-lived
quarks are formed dynamically in the $\hat{m}_{f} = 0$ case, chiral
symmetry must be broken.
 
The second line of (A.31) moreover illustrates a technical but important
point:
residues of the two invariant functions of $S^{[1,0]-1}_{F}$ at the
commmon $p^{2} = -\kappa^{2}_{1,2}$ pole cannot be chosen independently
but must be related in such a way as to combine into a single pole in
the matrix-valued variable $p \hspace{-0.2cm}/$,
\begin{equation}
\frac{{-p \hspace{-0.2cm}/} + \kappa_{1,2}}{p^{2} + \kappa^{2}_{1,2}} =
\frac{1}{{p \hspace{-0.2cm}/} + \kappa_{1,2}} \, .
\end{equation}
The reason is that by using unrelated residues one obtains, in general,
a propagator $S_{F}$ with {\it{three}} poles, and therefore ends
up in the subsequence describing stable quark particles that we do not
wish to study here.
 
\subsection{Fermion-vector three-point vertices.}
Color structure is strictly identical to that of the bare vertices:
\begin{equation}
\Gamma^{\mu, a}_{F, c, \alpha; {\overline{F}}, c', \beta} =
\left( \frac{1}{2} \lambda^{a} \right)_{cc'}
\Gamma^{\mu}_{F, \alpha; {\overline{F}}, \beta}  (p_{1}, k, -p_{2})
\end{equation}
($c,c'$, triplet-color indices; $\alpha, \beta$, Dirac-Spinor indices,
$k$, gluon four-momentum, and $p_{1} + k - p_{2} = 0$). Lorentz
(matrix-vector) structure of the $\Gamma^{\mu}_{\alpha \beta}$
amplitudes is then the same as for the QED vertex [32]. We therefore
mention only that a tensor decomposition suitable for our purpose,
\begin{equation}
\Gamma^{\mu}_{\alpha \beta} (p_{1}, k, - p_{2}) = \sum
\limits^{12}_{i=1} \, W_{i} \left(p^{2}_{1}, k^{2}, p^{2}_{2} \right)
\left(V^{\mu}_{i} \right)_{\alpha \beta} \, ,
\end{equation}
is possible in terms of the twelve matrix-valued vectors,
\begin{equation}
V_{1} = \gamma^{\mu}, \quad V_{2} = {p \hspace{-0.2cm}/}_{1}
\gamma^{\mu}, \quad V_{3} = \gamma^{\mu} {p \hspace{-0.2cm}/}_{2},
\quad V_{4} = {p \hspace{-0.2cm}/}_{1} \gamma^{\mu} {p
\hspace{-0.2cm}/}_{2} \, ,
\end{equation}
\begin{equation}
V_{5} = r^{\mu} {\rm{1\!I}}, \quad V_{6} = {p \hspace{-0.2cm}/}_{1}
r^{\mu}, \quad V_{7} = r^{\mu} {p \hspace{-0.2cm}/}_{2},
\quad V_{8} = {p \hspace{-0.2cm}/}_{1} r^{\mu} {p
\hspace{-0.2cm}/}_{2} \, ,
\end{equation}
\begin{equation}
V_{9} = k^{\mu} {\rm{1\!I}}, \quad V_{10} = {p \hspace{-0.2cm}/}_{1}
k^{\mu}, \quad V_{11} = k^{\mu} {p \hspace{-0.2cm}/}_{2},
\quad V_{12} = {p \hspace{-0.2cm}/}_{1} k^{\mu} {p
\hspace{-0.2cm}/}_{2} \, ,
\end{equation}
where
\begin{equation}
r^{\mu} = \frac{1}{2} (p_{2} + p_{1})^{\mu}, \quad k^{\mu} = (p_{2} -
p_{1})^{\mu} \, .
\end{equation}
(The vertices for {\it{transverse}} gluons, $\Gamma_{FT \overline{F}}$,
clearly contain only the first eight, eqs. (A.38/39), of these
matrices). In this basis, the restrictions imposed by $C$ invariance,
which leave nine of the twelve scalar functions $W_{i}$ independent,
take rather simple forms. In particular,
\begin{equation}
W_{i} \left( p^{2}_{1}, k^{2}, p^{2}_{2} \right) = W_{i+1} \left(
p^{2}_{2}, k^{2}, p^{2}_{1} \right) \quad (i = 2 \, \, {\rm{or}} \, \,
6) \, ,
\end{equation}
\begin{equation}
W_{i} \left( p^{2}_{1}, k^{2}, p^{2}_{2} \right) = W_{i} \left(
p^{2}_{2}, k^{2}, p^{2}_{1} \right) \quad (i = 1, 4, 5, 8)\, ,
\end{equation}
so that the transverse part has six independent invariant functions,
with four of these being symmetric under interchange of the fermion
and anti-fermion variables.
 
Construction of the FDRA sequence is greatly simplified by anticipating
that DS self-consistency will dictate, for the fermionic variables, the
correlation between pole and matrix structures indicated in (A.35),
leaving only "matrix-valued poles", with the ordering of matrices
adopted in (A.38 - 40) always observed. The resulting transverse,
nonperturbative vertex of zeroth perturbative order, obeying
conditions (2.3/2.4), will be written in the form of a partial-fraction
decomposition:
\begin{equation}
\left[ \Gamma^{\nu}_{FT {\overline{F}}} (p_{1}, k, -p_{2})
\right]^{[r,0]} = t^{\nu \mu} (k) \left\{ C^{\mu}_{0,r} (p_{1},-p_{2}) +
\sum \limits^{r}_{t=1} \, C^{\mu}_{t,r} (p_{1}, -p_{2}) \left(
\frac{\Lambda^{2}}{k^{2} + {\overline{u}}'_{r,2t} \Lambda^{2}} \right)
\right\} \, ,
\end{equation}
where
\begin{equation}
\begin{array}{rl}
C^{\mu}_{0,r} (p_{1}, - p_{2}) = & \gamma^{\mu} + \sum \limits^{r}_{s=1}
\, z^{(r)}_{0,1,s} \left[ \frac{\Lambda}{{p \hspace{-0.1cm}/}_{1} +
\kappa'_{r,2s}} \gamma^{\mu} + \gamma^{\mu} \frac{\Lambda}{{p
\hspace{-0.1cm}/}_{2} + \kappa'_{r,2s}} \right]\\[3ex]
\mbox{} & + \sum \limits^{r}_{s,s'=1} \, \frac{\Lambda}{{p
\hspace{-0.1cm}/}_{1} + \kappa'_{r,2s}} \left[ z^{(r)}_{0,4,s,s'}
\gamma^{\mu} + z^{(r)}_{0,5,s,s'} \left( \frac{r^{\mu}}{\Lambda} \right)
\right] \frac{\Lambda}{{p \hspace{-0.1cm}/}_{2} + \kappa'_{r,2s'}} \,
,\\[3ex]
\mbox{} & + \left[ {\rm{terms \, \, \, with\, \, \, }} (k^{2})^{n}\, ,
\,\, 1 \le n \le \frac{r-1}{2}\right]
\end{array}
\end{equation}
\mbox{}\\
\mbox{}\\
\begin{equation}
\begin{array}{rl}
C^{\mu}_{t,r} (p_{1}, -p_{2}) = & z^{(r)}_{t,0} \gamma^{\mu} +
z^{'(r)}_{t,0} \left( \frac{r^{\mu}}{\Lambda} \right) +
z^{''(r)}_{t,0} \left( \frac{{p \hspace{-0.1cm}/}_{1}}{\Lambda}
\gamma^{\mu} + \gamma^{\mu} \frac{{p \hspace{-0.1cm}/}_{2}}{\Lambda}
\right) + \\[3ex]
\mbox{} & + \sum \limits^{r}_{s=1} \, \frac{\Lambda}{{p
\hspace{-0.1cm}/}_{1} + \kappa'_{r,2s}} \left[ \left( z^{(r)}_{t,1,s}
+ z^{'(r)}_{t,1,s} \frac{p^{2}_{2}}{\Lambda^{2}}\right) \gamma^{\mu} +
z^{(r)}_{t,2,s} \gamma^{\mu} \frac{{p \hspace{-0.1cm}/}_{2}}{\Lambda}
+ z^{(r)}_{t,3,s} \frac{r^{\mu}}{\Lambda}\right.\\[3ex]
\mbox{} & \left. + z^{'(r)}_{t,3,s} \frac{r^{\mu} {p
\hspace{-0.1cm}/}_{2}}{\Lambda^{2}} \right]  + \sum \limits^{r}_{s=1} \,
\left[ \left( z^{(r)}_{t,1,s} + z^{'(r)}_{t,1,s}
\frac{p^{2}_{1}}{\Lambda^{2}} \right) \gamma^{\mu} + z^{(r)}_{t,2,s}
\frac{{p \hspace{-0.1cm}/}_{1}}{\Lambda} \gamma^{\mu}\right.\\[3ex]
\mbox{} & \left. + z^{(r)}_{t,3,s} \frac{r^{\mu}}{\Lambda} +
z^{'(r)}_{t,3,s}
\frac{{p \hspace{-0.1cm}/}_{1} r^{\mu}}{\Lambda^{2}} \right]
\frac{\Lambda}{{p \hspace{-0.1cm}/}_{2} + \kappa'_{r,2s}} + \sum
\limits^{r}_{s,s'=1} \, \frac{\Lambda}{{p \hspace{-0.1cm} /}_{1} +
\kappa'_{r,2s}} \times\\[3ex]
\mbox{} & \times \left[ z^{(r)}_{t,4,s,s'} \gamma^{\mu} +
z^{(r)}_{t,5,s,s'} \frac{r^{\mu}}{\Lambda} \right] \frac{\Lambda}
{{p \hspace{-0.1cm}/}_{2} + \kappa'_{r,2s'}} \, .
\end{array}
\end{equation}
Again, preservation of perturbative renormalizability turns out to be a
requirement slightly stronger than (2.3/2.4), imposing extra
restrictions in the case where $p_{1}, p_{2}$ run to infinity in a loop
while $k$ stays outside the loop and fixed. For the vertex to behave
no worse than a constant at large loop momenta in this situation
requires the primed coefficients in (A.46), or sums of these, to vanish:
\begin{equation}
z^{'(r)}_{t,0} \quad = \quad z^{''(r)}_{t,0} \quad = \quad 0
\qquad ( 1 \le t \le r, \,\, {\rm{all}} \,\, r ) \, ;
\end{equation}
\begin{equation}
\sum \limits_{s=1}^{r} \, z^{'(r)}_{t,1,s} \quad = \quad \sum \limits
_{s=1}^{r} \, z^{'(r)}_{t,3,s} \quad = \quad 0
\qquad ( 1 \le t \le r, \,\, {\rm{all}} \,\, r ) \, .
\end{equation}
 
In the remaining terms, mass scales have for simplicity been written
in a form appropriate for the chiral limit, $\hat{m}_{f} = 0$. At
$\hat{m}_{f} \not= 0$, those mass scales in (A.45/46) that are not
forced by condition (2.3) to be pure multiples of $\Lambda$ should be
generalized in an obvious way, following the pattern of (A.28/29).
Formally, this can be accounted for by allowing for a dependence on the
dimensionless ratio $\hat{m}_{f}/\Lambda$ in the relevant dimensionless
$z$ coefficients.
In fact, in sufficiently high loop orders, {\it{all}}
nonperturbative coefficients, including those of the gluon-ghost sector,
will eventually depend on {\it{all}} $\hat{m}_{F}/\Lambda$ ratios.
\end{appendix}
\newpage
\begin{center}
{\large \bf{Figure Captions}}
\end{center}
\bigskip
\bigskip
\begin{description}
\item[Fig. 1] Distribution of poles (crosses) and zeroes (circles) for
nonperturbatively modified propagation functions $D^{[r,0]}_{T} (k^{2})$
in (A) "particle" subsequence, (B) "quasiparticle" subsequence.
\bigskip
\item[Fig. 2] Diagrammatic form of Dyson-Schwinger equation for
two-point vertex (negative-inverse propagator) of gluon field,
showing one-DS-loop terms (A $\ldots$ D) and two-DS-loops terms
$(E \& F)$ of dressing functional.
\bigskip
\item[Fig. 3] One-particle-reducible contributions to four-point
amplitudes of elementary QCD fields that seem to produce causality
violation in S-matrix elements for the "quasiparticle" subsequence.
\bigskip
\item[Fig. 4] Qualitative behavior of propagation function $\tilde{D}$
in Euclidean coordinate space for free massive propagator (narrow line)
and a "quasiparticle" propagator of type $\tilde{D}^{[1,0]}$
(broad line)
\bigskip
\item[Fig. 5] Schematic view of probability flow through state space
of (I) an ordinary scattering system and (II) an asymptotically
incomplete system with a closed subspace ${\cal{H}}_{C}$.
\end{description}
\newpage
\epsfxsize=15cm \epsfbox{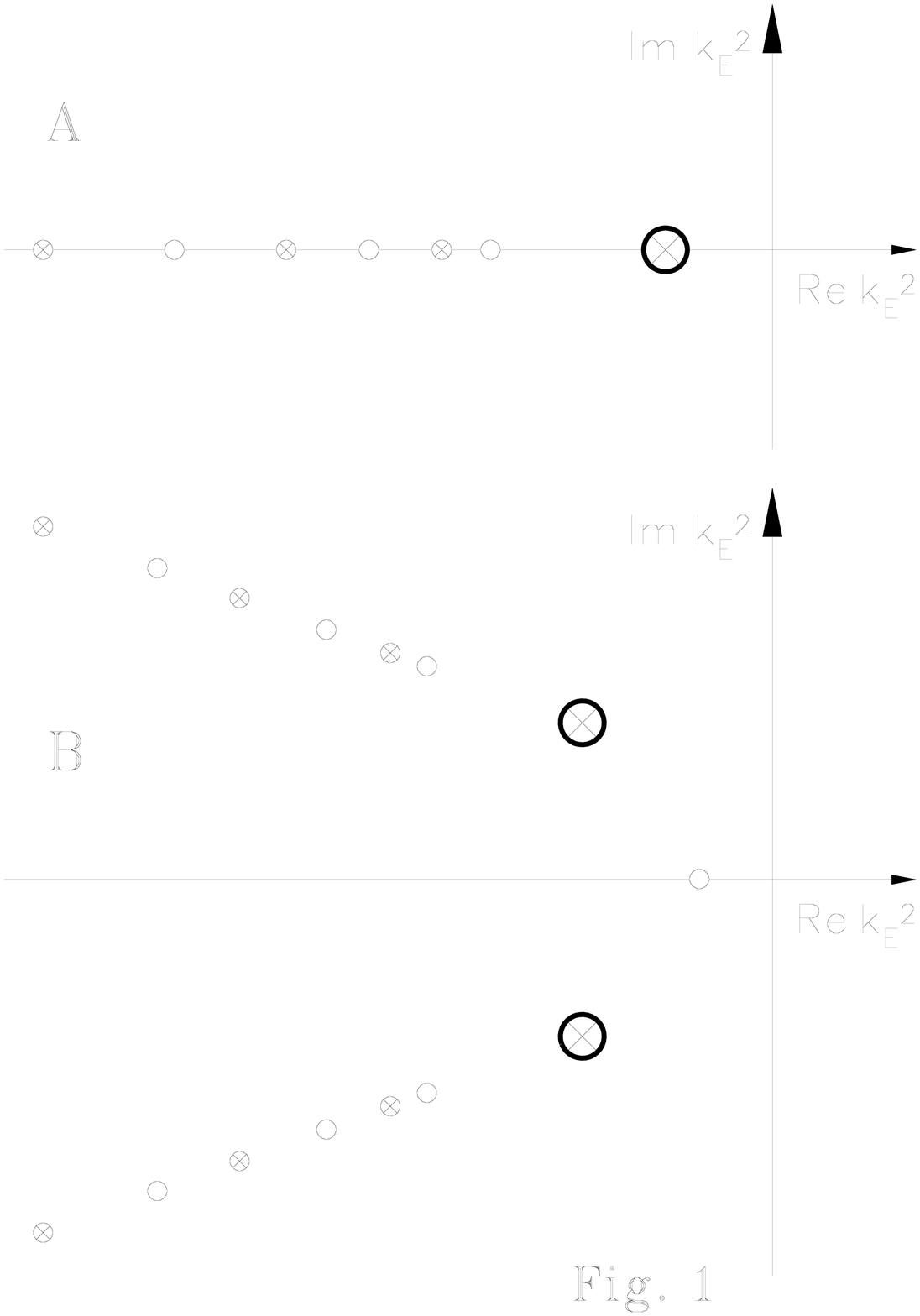}
\newpage
\epsfxsize=15cm \epsfbox{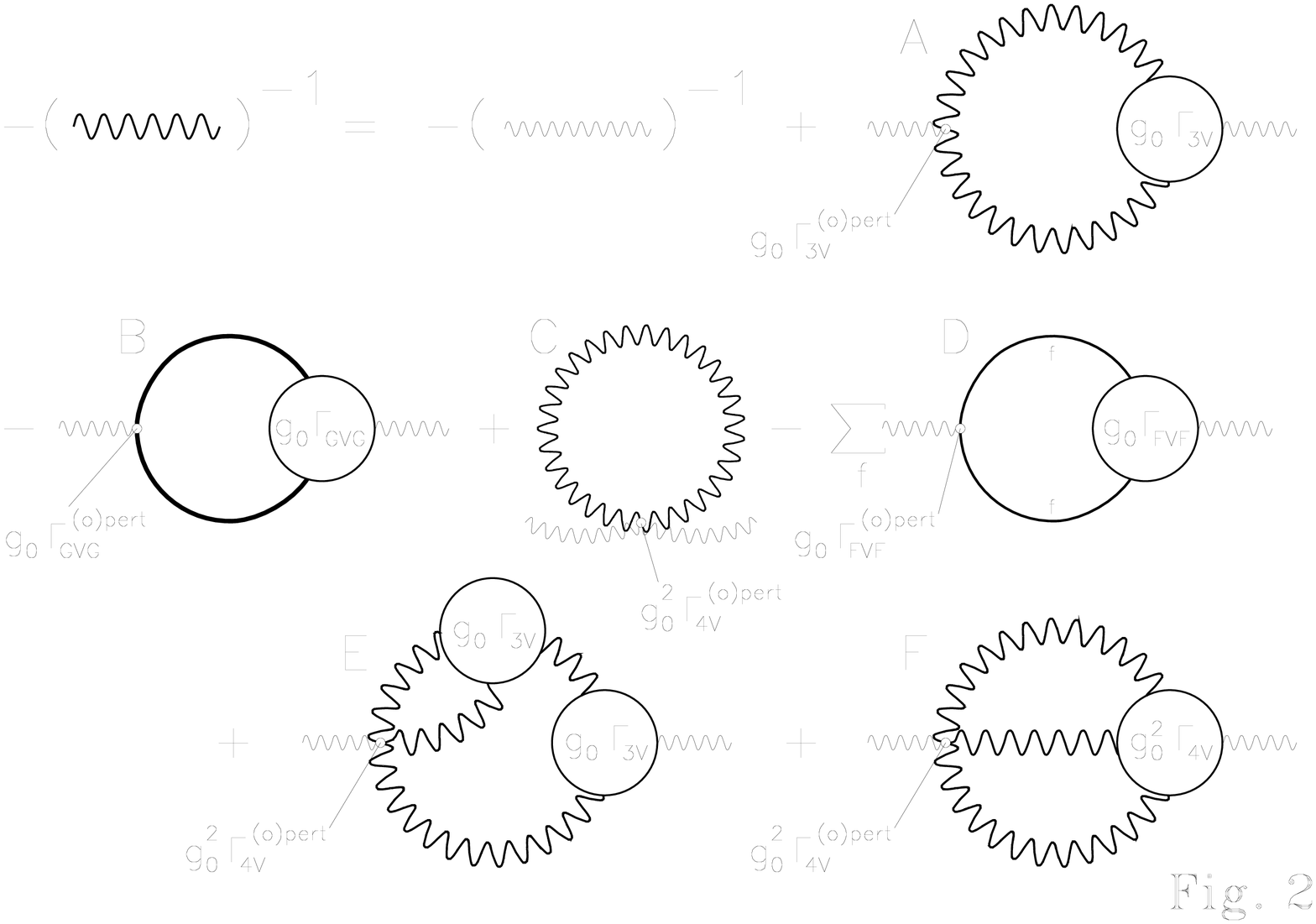}

\vspace{2cm}

\epsfxsize=15cm \epsfbox{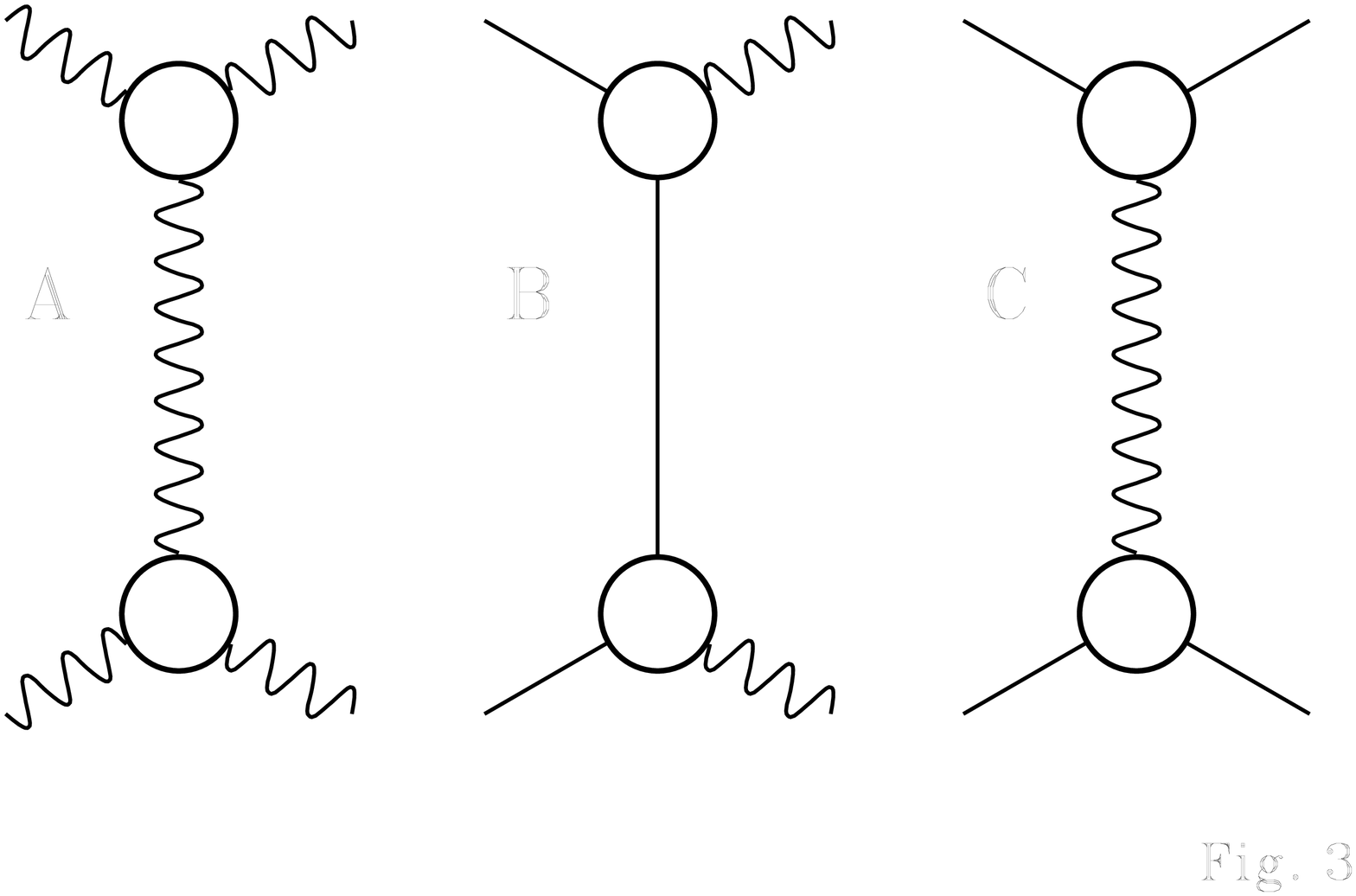}
\newpage
\epsfxsize=15cm \epsfbox{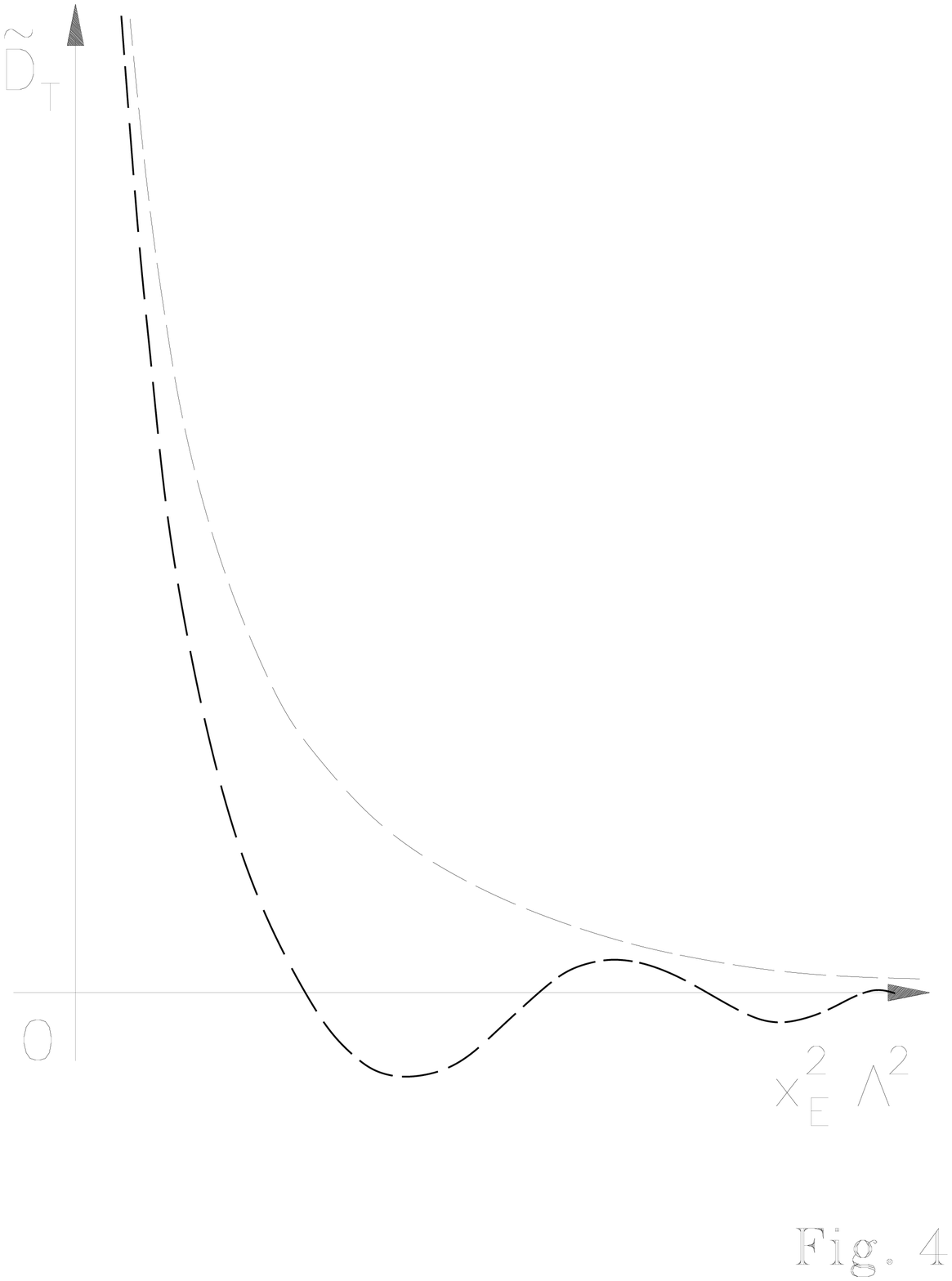}
\newpage
\epsfxsize=15cm \epsfbox{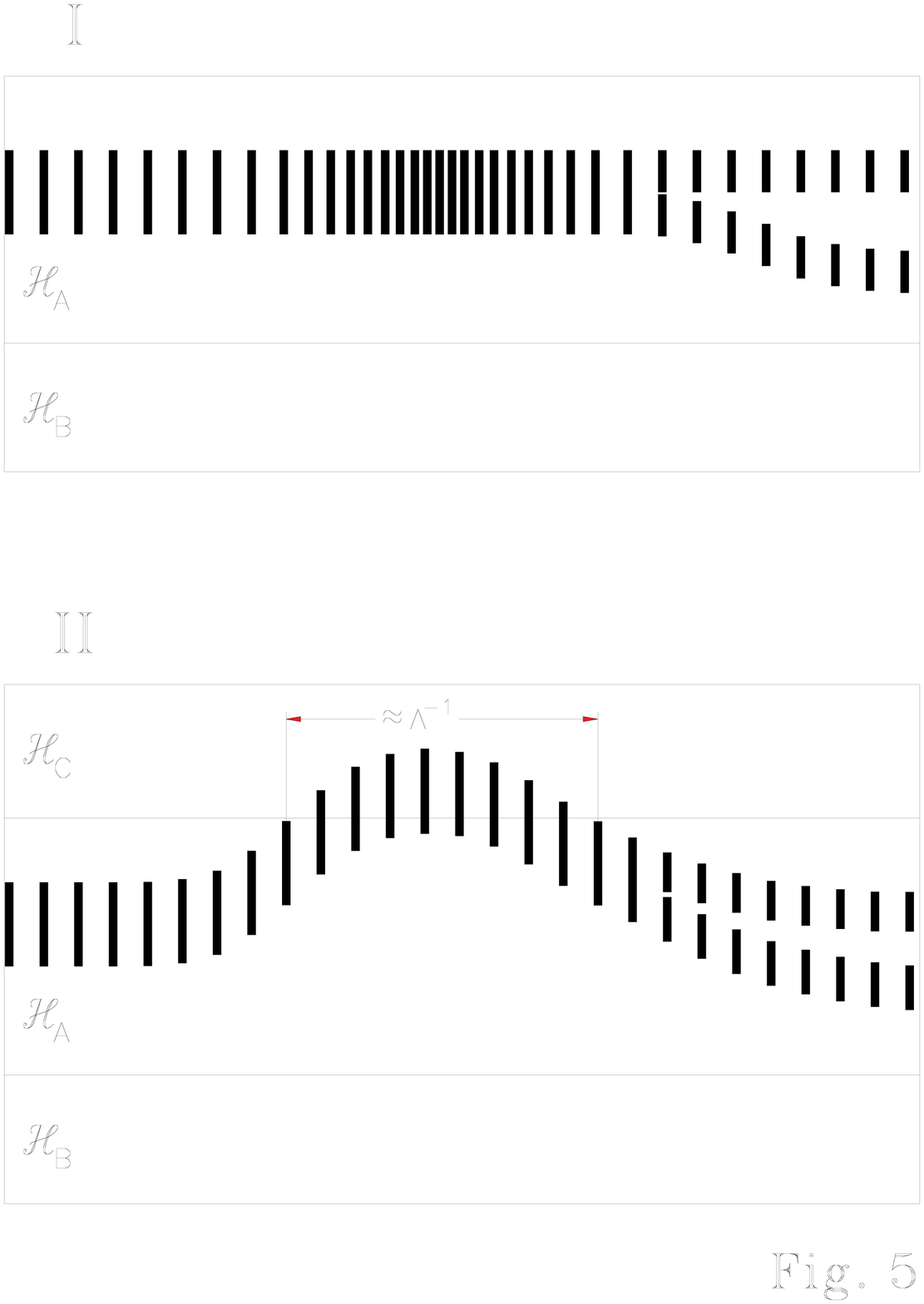}
\end{document}